\documentclass[onecolumn,authoryear]{els-mrw} 

\usepackage{amsmath,amssymb,amsfonts,amsthm,makeidx,graphicx}
\usepackage{txfonts}
\usepackage{helvet}

\usepackage{hyperref}        

\bibpunct{(}{)}{;}{}{}{,}

\def\gtsim {>\kern-1.2em\lower1.1ex\hbox{$\sim$}~}   
\def\ltsim {<\kern-1.2em\lower1.1ex\hbox{$\sim$}~}   

\begin{document}

\chapter{Nucleosynthesis and the chemical enrichment of galaxies}\label{chap1}

\author[1]{Chiaki Kobayashi}

\address[1]{\orgname{Centre for Astrophysics Research}, \orgdiv{Department of Physics, Astronomy and Mathematics, 
University of Hertfordshire}, \orgaddress{College Lane, Hatfield  AL10 9AB, UK }}

\articletag{Chapter Article tagline: update of previous edition,, reprint..}

\maketitle

\begin{abstract}[Abstract]
Stars are fossils that retain the history of their host galaxies. Carbon and heavier elements are created inside stars and are ejected when they die. From the spatial distribution of elements in galaxies, it is therefore possible to constrain the physical processes during galaxy formation and evolution. 
This approach, Galactic archaeology, has been popularly used for our Milky Way Galaxy thanks to a vast amount of data from the Gaia satellite and multi-object spectrographs, and now can also be applied to very distant galaxies with the James Webb Space Telescope (JWST) -- extra-galactic archaeology.
In these studies the most important factor is the input stellar physics, namely nucleosynthesis yields and binary physics, which predominantly determine the model predictions.
In this review I give a summary of stellar nucleosynthesis, and how they are tested with the observations in the Milky Way. Then I show how chemical enrichment of galaxies can be calculated, and show some results with the latest nucleosynthesis yields.
\end{abstract}

\section*{Key Points}
\begin{itemize}
\item {\bf Nucleosynthesis yields} are the mass, or mass fraction, of elements and isotopes from stars, calculated in stellar evolution and supernova explosions based on nuclear physics.
\item {\bf $\alpha$ elements} (O, Ne, Mg, Si, S, Ar, and Ca) are mainly produced by core-collapse supernovae, more from more massive ones.
\item {\bf Iron-peak elements} are largely from Type Ia supernovae (high Mn/Fe with high-mass explosion) as well as hypernovae (high (Co,Zn)/Fe with high energy).
\item {\bf The initial mass function} (IMF) gives the number, or mass contribution, of stars formed at a given initial mass.
\item {\bf The star formation history} (SFH) is time evolution of the star formation rate.
\item {\bf The [$\alpha$/Fe]--[Fe/H] relation} is caused by the delayed Fe enrichment; [$\alpha$/Fe] ratio is roughly constant (with a value depending on IMF) at low metallicities, and sharply decreases at [Fe/H] $\sim -1$ in the solar neighborhood. The `knee' depends on SFH; rapid star formation in e.g. the bulge and massive galaxies causes the [$\alpha$/Fe] decrease at a higher [Fe/H].
\item {\bf The N/O--O/H relation} is caused by the primary process with mixing/rotation in massive stars, the primary process in asymptotic giant branch (AGB) stars, and the secondary process in all mass ranges of stars. The last two causes an increase of N/O at high metallicities, depending on IMF and SFH.
\item {\bf Odd-Z elements} (e.g., Na, Al, Mn, Cu) and minor isotope yields depend on the initial N abundances, and thus on the initial metallicity and stellar rotation, of progenitor stars.
\item {\bf The first peak neutron-capture elements} (Sr, Y, Zr) are produced from AGB stars, electron-capture supernovae, and possibly rotating massive stars.
\item {\bf The slow neutron-capture process} occurs in AGB stars, enhancing Ba and Pb.
\item {\bf The rapid neutron-capture process} occurs in neutron star mergers (NSM), magneto-rotational supernovae (MRSN), and maybe collapsars, enhancing Eu. It is still a question which site(s) produce actinides.
\item {\bf Extremely metal-poor (EMP) stars} have [Fe/H] $<-3$ and are likely to be formed in the early Universe, possibly enriched by only one or a few supernovae.
\item {\bf Carbon-enhanced metal-poor (CEMP) stars} have [C/Fe] $>0.7$, and many of EMP stars are CEMP stars. Those with and without s-process enhancement are called CEMP-s and CEMP-no, respectively.
\item {\bf Galactic chemical evolution} (GCE) models predict the evolution of elemental abundances and isotopic ratios, assuming IMF, SFH and gas flows (inflow, outflow, and radial flow) of the galaxy.
\item {\bf The metallicity distribution function} (MDF) gives the number of stars at a given metallicity, and offers a strong constraint for GCE modelling, but is available only in the Local Group galaxies.
\item {\bf The mass--metallicity relation} (MZR) of integrated stellar populations or the inter-stellar medium (ISM) can be used for other galaxies, and it is primarily caused by the metal-loss in low-mass galaxies.
\item {\bf Metallicity radial gradients} are a simple description for the spatial distribution of elements in a galaxy, and can constrain SFH and gas flows within the galaxy.
\item {\bf Chemo-hydrodynamical simulations} can predict the spatial distributions of elements and isotopes within the galaxy including cosmological growth, star formation and gas flows.
\end{itemize}

\section{Introduction}

\begin{figure}[t]
\begin{center}
\includegraphics[width=8cm]{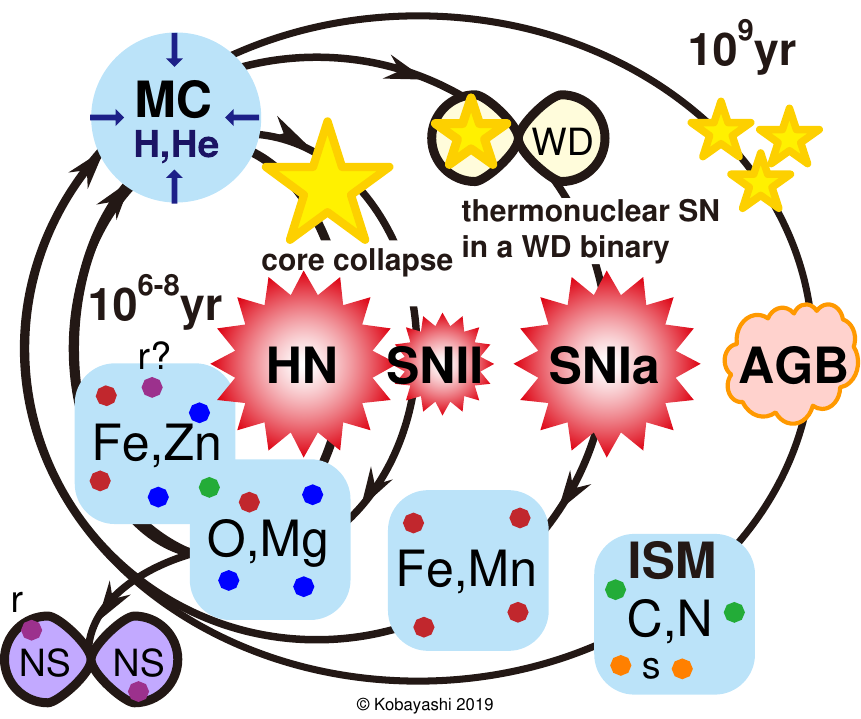}
\caption{\label{fig:intro}
Schematic view of chemical enrichment in galaxies from hypernova (HN), Type II supernova (SN II), Type Ia supernova (SN Ia), asymptotic giant branch star (AGB), and neutron star (NS) merger.
}
\end{center}
\end{figure}

Explaining the origin of the elements is one of the scientific triumphs linking nuclear physics with astrophysics. As Fred Hoyle predicted, carbon and heavier elements (`metals' in astrophysics) were not produced during the Big Bang but instead created inside stars. So-called $\alpha$ elements (O, Ne, Mg, Si, S, Ar, and Ca) are mainly produced by core-collapse supernovae, while iron-peak elements (Cr, Mn, Fe, and Ni) are more produced by thermonuclear explosions, observed as Type Ia supernovae (SNe Ia; \citealt{kob20sr}, hereafter K20). The element production depends on the mass of white dwarf (WD) progenitors, and a large fraction of SNe Ia should come from near-Chandrasekhar (Ch) mass explosions (see \citealt{kob20ia} for constraining the relative contribution between near-Ch and sub-Ch mass SNe Ia).
Among core-collapse supernovae, hypernovae ($\gtsim 10^{52}$ erg) produce a significant amount of Fe as well as Co and Zn, and a significant fraction of massive stars ($\gtsim 20M_\odot$) should explode as hypernovae in order to explain the Galactic chemical evolution (GCE; \citealt{kob06}).

Heavier elements than Fe are produced by neutron-capture processes. The slow neutron-capture process (s-process) occurs in asymptotic giant branch (AGB) stars \citep[e.g.,][]{busso99,her05,cri11,kar14}, while the astronomical sites of rapid neutron-capture process (r-process) have been debated. The possible sites are neutron-star (NS) mergers \citep[NSMs,][]{wan14,jus15}, magneto-rotational supernovae \citep[MRSNe,][]{nis15,mos18,rei21} or hypernovae \citep[MRHNe,][]{yon21a}, 
accretion disks/collapsars (\citealt{sie19}, but see \citealt{just22}),
and common envelope jets supernovae \citep{gri22}. Light neutron-capture elements (e.g., Sr, Y, Zr) are also produced by electron-capture supernovae (ECSNe, \citealt{wan13ec}), $\nu$-driven winds \citep[weak-r,][]{arc07,wan13nu}, and rotating-massive stars \citep[weak-s,][]{fri16,lim18}.

The cycles of chemical enrichment are schematically shown in Figure \ref{fig:intro}, where each cycle produces different elements and isotopes with different timescales.
In a galaxy, not only the total amount of metals, i.e. metallicity $Z\equiv\Sigma_{i\ge{\rm C}}\,m_i/m$, but also elemental abundance ratios, [X/Fe]$\equiv{\log N_X/N_{\rm Fe}/(N_{X,\odot}/N_{{\rm Fe},\odot}})$, evolve as a function of time. 
Therefore, we can use all of this information as fossils to study the formation and evolutionary histories of the galaxy. This approach is called Galactic archaeology, and several on-going and future surveys with multi-object spectrographs (MOS; e.g., APOGEE, HERMES-GALAH, Gaia-ESO, DESI, WEAVE, 4MOST, MOONS, Subaru Prime Focus Spectrograph (PFS), 
High-Resolution Multi-Object Spectrograph (HRMOS) for VLT, 
and Wide-field Spectroscopic Telescope (WST)) 
are producing a vast amount of observational data of elemental abundances.
Moreover, integral field unit (IFU) spectrographs (e.g., SAURON, SINFONI, CALIFA, SAMI, MaNGA, KMOS, MUSE, HECTOR, and NIRSpec on JWST) allow us to measure metallicity and some elemental abundance ratios within galaxies. It is now possible to apply the same approach not only to our own Milky Way but also to other types of galaxies or distant galaxies. Let us call this extra-galactic archaeology.

\begin{figure}[t]
\begin{center}
\includegraphics[width=0.6\textwidth]{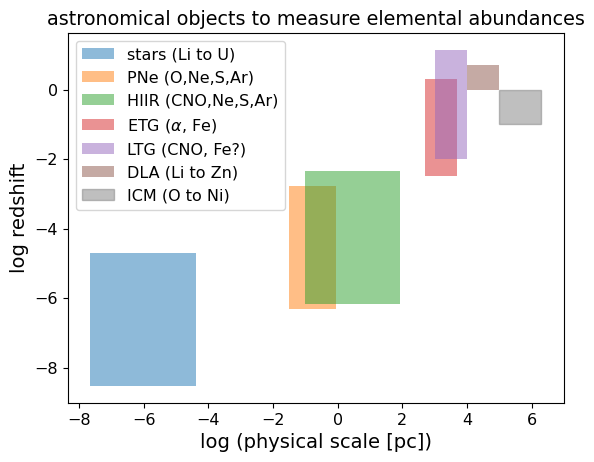}
\caption{\label{fig:obs}
Astronomical objects with elemental abundance measurements: individual stars in the local Universe (blue), planetary nebulae (orange), HII regions (green), stellar populations in early-type galaxies (red), the interstellar medium in late-type star-forming galaxies (purple), the damped Lyman $\alpha$ systems (brown), and the intra-cluster medium (gray).
}
\end{center}
\end{figure}

Elemental abundances can be measured from observed spectra of stars or galaxies, as atoms emit or absorb light at specific wavelengths from transitions between electron energy levels, according to atomic physics.
Various astronomical observations are summarized in Figure \ref{fig:obs}.
Absorption lines in high-resolution ($R>50,000$) spectra of the Sun and nearby stars provide the most detailed elemental abundances from Li to U (blue), very accurately if the effects of the three-dimensional (3D) and non-local thermal equilibrium (NLTE) in the stellar atmosphere are modelled \citep{lind24}. With emission lines in planetary nebulae (PNe; orange) or HII regions (HIIR; green), it is possible to measure some elements (e.g., CNO, Ne, S, Ar) in neighbour galaxies e.g., the Andromeda Galaxy \citep[e.g.,][]{kob23}.
But for other external galaxies, it is no longer possible to resolve individual stars, and thus the integrated spectra of stellar population or the interstellar medium (ISM) in lower resolution spectra have been used. From stellar absorption lines of early-type galaxies (ETG; red), $\alpha$ enhancement relative Fe has been widely used to constrain galaxy formation \citep{tho03}, followed by the attempts to estimate several more elements \citep{conroy13}. Nebular emissions from the ISM powered by stars in late-type galaxies (LTG; purple) can provide several elemental abundances \citep{kew19,mai19}. Due to ``redshift'' of spectra, this was available only for $z\sim2$, but followed by the extension by the Atacama Large Millimeter/Submillimeter Array (ALMA) \citep{harikane20,franco21}, the JWST greatly extended the redshift range  \citep{bunker23,deugenio24,carnall24}.
In studies of cosmic chemical evolution, absorption lines along the line of sights of quasars (or $\gamma$-ray bursts) have also been used, and the damped Ly$\alpha$ systems (DLA; brown) provide the most accurate measurements of elemental abundances of gas at high-redshifts \citep{pettini04,wolfe05}.
Finally, although only in the local Universe, X-ray hot gas in intra-cluster medium (ICM; gray) also provide detailed elemental abundances \citep[e.g.,][]{hitomi17}.

While the evolution of the dark matter in the standard $\Lambda$ cold dark matter (CDM) cosmology is reasonably well understood, how galaxies form and evolve is still much less certain because of the complexity of the baryon physics such as star formation and feedback. The thermal energy ejected from supernovae to the ISM suppresses star formation, while the production of heavy elements in these supernovae enhances gas cooling. Since these processes affect each other, galaxy formation and evolution is complicated, and has to be solved consistently with a numerical simulation. Feedback from central super-massive black holes (SMBHs) is also found to be very important for explaining the observed properties of galaxies \citep[e.g.,][]{cro06,spr05agn,hop06,dub12,tay14}.

Since the 90’s, the development of high performance computers and computational techniques has made it possible to simulate the formation and evolutionary history of galaxies, not only of isolated systems \citep[e.g.,][]{bur87,kat92,ste94,mih96,kaw03,kob04} but also for cosmological simulations of individual galaxies
\citep[e.g.,][]{nav94} or of the galaxy population as a whole \citep[e.g.,][]{cen99,kob07}.
Thanks to the public release of hydrodynamical simulation codes such as Gadget \citep{spr01}, RAMSES \citep{tey02}, and AREPO \citep{spr10}, it became easier to run galaxy simulations, and even simulation data are made public (e.g., EAGLE, Illustris).
However, to cover a wide range of galaxy mass in cosmological `box' simulations, the simulation volume has to be sufficiently large, which limits numerical resolutions.
On the other hand, `zoom-in' techniques allow us to increase the resolution focusing a particular galaxy,
but this also requires tuning the parameters with the same resolution, comparing to a number of observations in the Milky Way \citep{sca12}.

Needless to say, these simulation results highly depend on the input baryon physics, namely nuclear astrophysics, for predicting chemical abundances.
In this review I give a summary of stellar nucleosynthesis, and how the yields are constrained with observations using a simple chemical evolution model in the Milky Way. Then I show how chemical enrichment can be calculated in more sophisticated, chemodynamical simulations of galaxies.
See \citet{kob23book} for a review on chemodynamical simulations including formula.

\section{Nucleosynthesis yields}

\begin{figure}[t]
\begin{center}
\includegraphics[width=0.6\textwidth]{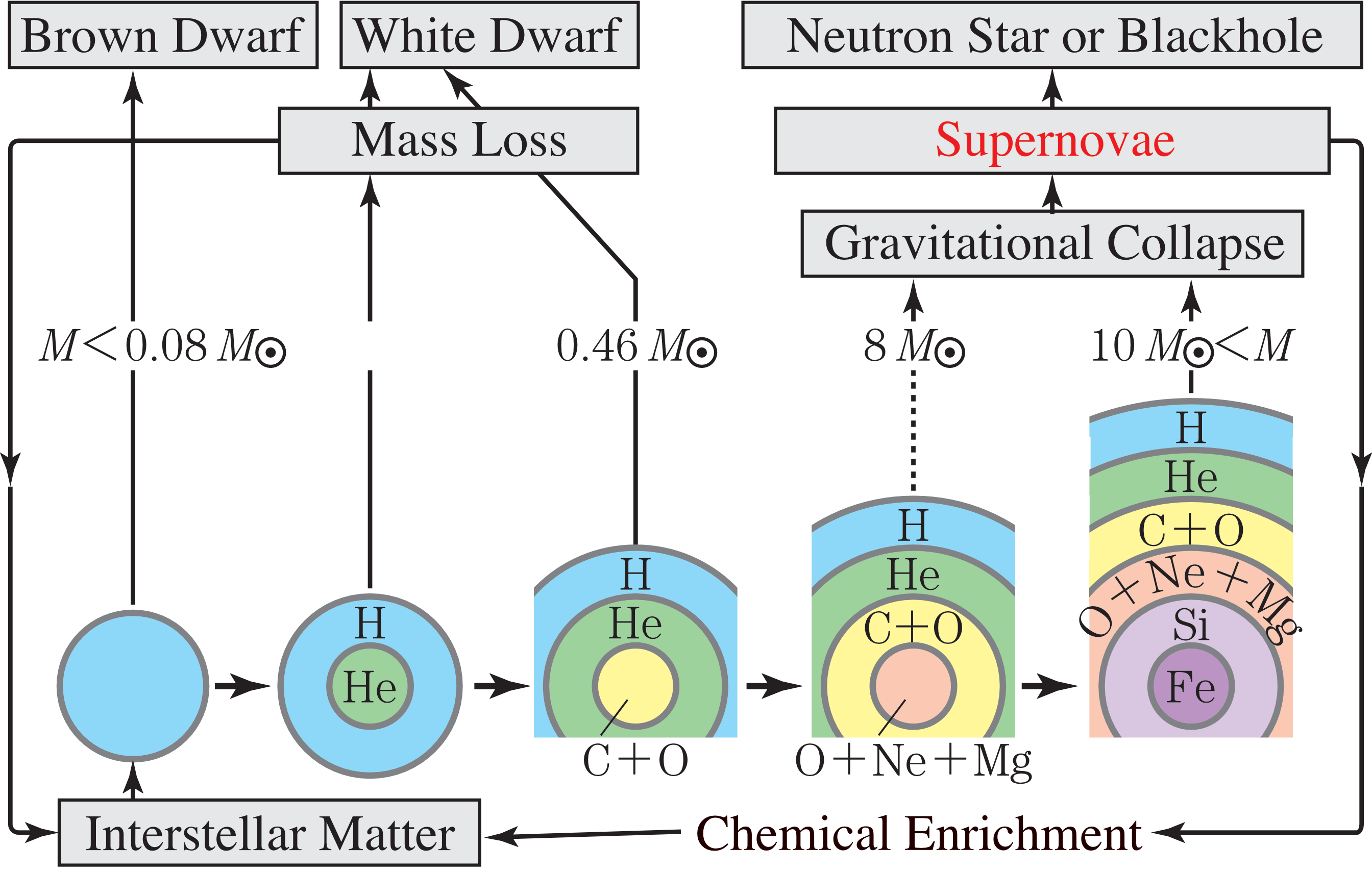}
\caption{\label{fig:fate}
The fate of stars depending on the initial mass $M$ (in the case of single stars). The threshold masses depend on the metallicity and are for $Z=Z_\odot$.
[Credit: K. Nomoto 2007, Iwanami-shoten]
}
\end{center}
\end{figure}

During the Big Bang, only light elements are produced because of the lack of stable nuclei with the mass number $A=5$ and 8 \citep[for more details]{coc12}.
Carbon is formed by the triple-$\alpha$ reactions (3 $^4$He $\rightarrow$ $^{12}$C $+\gamma$), for which \citet{hoyle54} predicted a resonance state, at high density and high temperature i.e. in stars.
Stars shine thanks to the nuclear energy transforming light nuclei into heavier but more stable nuclei. 
As a protostar contracts, its central temperature increases, and when it reaches $\sim 10^7$ K, hydrogen fusion occurs via the p-p chain, then the CNO cycle. In the Sun, 98\% of energy is produced by the p-p chain.
After $\sim$10 billion years from the birth, stars like our Sun form C+O cores, but C-ignition occurs only in intermediate-mass stars off-centre (with initial masses of $\sim$8--9$M_\odot$) or at the centre ($\gtsim9M_\odot$, \citealt{kob15}).
If Ne-burning is not ignited or if off-centre Ne-burning does not propagate to the centre, the O+Ne+Mg core may collapse as an electron-capture supernova (ECSN) depending on the mass loss,
which leaves a neutron star (NS) as observed in the Crab Nebula -- the supernova remnant of SN1054
\citep{nom13}.
Stars ten times more massive than the Sun ($\gtsim10M_\odot$) form Fe cores.
Iron is one of the most stable elements with a large binding energy per nucleon, and thus it is not easy to synthesize much heavier elements than iron.
Instead, due to photodissociation of iron ($^{56}$Fe $\rightarrow$ 13$^{4}$He$+4$n$-\gamma$), gravitational collapse is triggered at the centre of massive stars.
Although the remnant formation is highly uncertain, possibly, stars $\ltsim25M_\odot$ may form a neutron star (NS), while stars $\gtsim25M_\odot$ may form a black hole (BH).
Recently, the JWST observation confirmed the existence of a neutron stars in SN1987A \citep{fransson24}.
The fate of stars are summarized in Figure \ref{fig:fate}.
It is important to note that this figure is for single stars, and can be very different for binary stars due to binary interaction and mass transfer.

\subsection{Massive stars}

Pre-supernova yields are determined from these hydrostatic burning processes. The uncertainties include the reaction rates (namely, of $^{12}$C($\alpha$, $\gamma$)$^{16}$O), mixing in stellar interiors, rotationally induced mixing processes \citep{maeder00}, and mass loss via stellar winds \citep{vink22}. Although a short time range, three-dimensional hydrodynamical simulations of stellar evolution is now available, which could remove ad hoc procedures of convective overshooting and semi-convection \citep{arnett19}.
It is not possible to simulate the whole evolution of a star in 3D, and the 3D effects should be implemented in 1D stellar evolution codes by somehow, which might significantly affect the yields of some odd-Z elements (e.g., Cl, K, Sc).

Generally speaking, more massive stars have more massive envelopes, and have higher $\alpha$ element yields (O, Ne, Mg, Si, S, Ar, Ca), while the mass dependence of Fe yields is weak (see Figs.\,1--4 of \citealt{kob06}). 
Therefore, [$\alpha$/Fe] is lower for low-mass supernovae (13--15 $M_\odot$).
The low-$\alpha$ abundances seen for a small number of extremely metal-poor stars are likely caused by the mass dependence of [$\alpha$/Fe] \citep[\S \ref{sec:first}]{kob14}.
The [$\alpha$/Fe] mass-dependence also means that a top-heavy initial mass function (IMF) gives higher [$\alpha$/Fe] ratios, which might help to reproduce high [$\alpha$/Fe] ratios seen for massive early-type galaxies \citep[e.g.,][\S \ref{sec:cosmo}]{tho05,tay15}.

The yields of odd-$Z$ elements (Na, Al, P, ... and Cu) highly depend on the initial metallicity of the progenitor stars, as their production depends on the surplus of neutrons from $^{22}$Ne, which is made during He-burning from $^{14}$N produced in the CNO cycle (see Fig.\,5 of \citealt{kob06}).
It is obvious that these elements are useful for identifying accreted stars in the Galactic archaeology.
Stars formed in a small satellite galaxy before they merge tend to have low [(Na,Al)/Fe] ratios, which is indeed the case for the thick disk stars in our chemodynamical simulations of a Milky Way-type galaxy (\citealt{kob11mw}; \S \ref{sec:mw}).

\subsection{Core-collapse supernovae}
\label{sec:sn}
Explosive nucleosynthesis during a supernova explosion changes the elemental abundance distribution (see Fig.\,1 of \citealt{nom13}).
The explosion mechanism of core-collapse (Type II, Ib, and Ic) supernovae is uncertain although a few groups have presented multi-dimensional simulations of exploding 10--25$M_\odot$ stars \citep{mar09,kot12,jan12,bru13,bur21} with detailed neutrino transport.
This may be consistent with the lack of observed progenitors at supernova locations in the HST data \citep{sma09}.
Supernovae are bright thanks to thermal/internal energy deposited by the supernova shock wave, as well as radioactive decays of nuclei created in the explosion (e.g.,\,$^{56}$Ni).
The explosion energy and ejected Fe mass can be estimated from the observations of nearby supernovae, namely, the light-curves and spectra \citep[see Fig.2 of ][]{nom13}. 
At $\ltsim20M_\odot$ including SN1987A,
the ejected energy is found to be roughly constant, $10^{51}$ erg (which became the unit fifty-one-erg (foe)). However, as for SN1998bw associated with $\gamma$-ray burst GRB980425 \citep{iwa98}, more massive stars eject more energy and Fe, which are called `hypernovae'.
The explosion mechanism of hypernovae is particularly uncertain, but might be related to rotation and magnetic fields (see Table 2 of K20).
The rest of stars above $25 M_\odot$ seem to end with `failed' supernovae, not producing Fe at all.
Note that these are different from `faint' supernovae, which are observationally driven and are proposed as explanation for carbon-enhanced metal poor stars (CEMP; \S \ref{sec:first}).

These explosion simulations are computationally expensive and even with a post-process nucleosynthesis it is not possible to provide a set of nucleosynthesis yields varying mass and metallicities as required for GCE calculations. Instead, nucleosynthesis yields are provided by 1D calculations by three different groups (\citealt{woo95,nom97,lim03}, see Fig.\,5 of \citealt{nom13} for the comparison), and are constantly being updated (e.g., \citealt{heg10,lim18,kob20sr}; see also \citealt{rit18a}).

\begin{figure}[t]
\begin{center}
\includegraphics[width=0.6\textwidth]{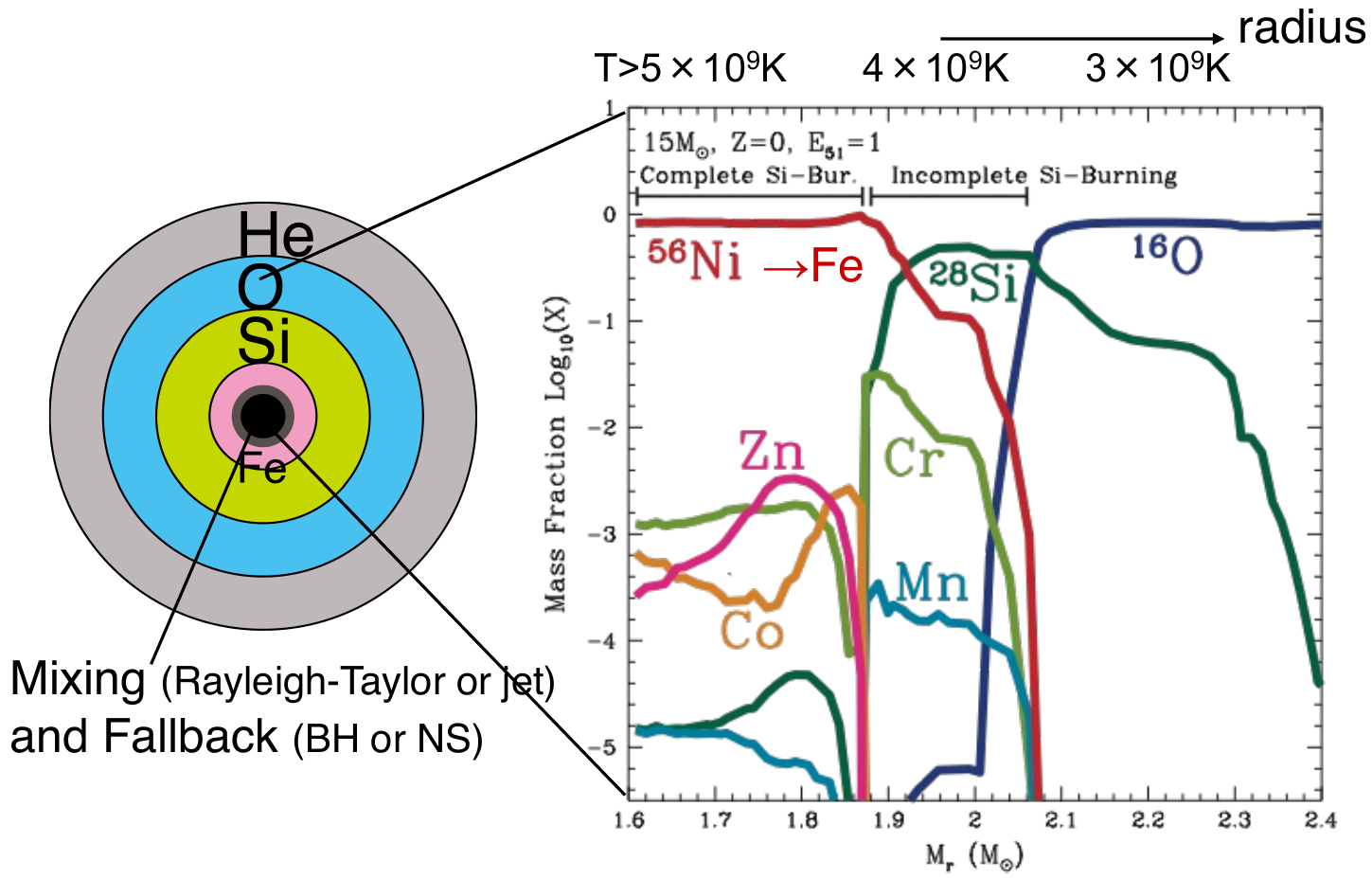}
\caption{\label{fig:snnuc}
Example of supernova nucleosynthesis of a massive star with $>10M_\odot$.
Abundance distribution after explosive burning is shown against an enclosed mass $M_r$ (right). 
}
\end{center}
\end{figure}

Figure \ref{fig:snnuc} shows a schematic view of supernova nucleosynthesis. After hydrostatic burning of  a massive star ($>10M_\odot$),  explosion energy is injected to follow the explosive nucleosynthesis.
Among Fe-peak elements, Cr and Mn are produced in the incomplete Si burning region, while Zn and Co are produced in the complete Si burning region.
The most central part (black) falls back to the remnant (NS or BH; mass-cut), while the layer around it (gray) is only partially ejected in our mixing-fallback model.
This ``mixing'' may be caused through Rayleigh–Taylor instability, or may be the result of jet-like explosion (see Fig.\,12 of \citealt{tom07}).
\citet{lim18} used the mass-cut only, while \citet{heg10} used a different way of mixing.

Our yields were originally calculated in \citet{kob06}, 3 models of which are replaced in \citet{kob11agb} (the identical table was also used in \citealt{nom13}), and a new set with failed supernovae is used in K20.
This resulted in a 20\% reduction of both the net and oxygen yields (see Table 3 of K20).
Supernova yields assumed $E_{51}\equiv E/10^{51} {\rm ergs}=1$ foe, while hypernova yields included the observed mass dependence of the explosion energy; $E_{51}=$ 10, 10, 20, 30 foe for 20, 25, 30, $40M_\odot$ stars.
These supernova and hypernova yield tables are provided separately, and it is recommended to assume $\epsilon_{\rm HN}=$50\% hypernova fraction for stars with $\ge 20M_\odot$ \citep{kob06}. This fraction is expected to decrease at high metallicities due to smaller angular momentum loss, and was assumed as $\epsilon_{\rm HN}(Z)=0.5, 0.5, 0.4, 0.01$, and $0.01$ for $Z=0, 0.001, 0.004, 0.02$ in \citet{kob11mw}.
The mass threshold of failed supernovae is determined to be $30M_\odot$ from Fig.\,4 of K20, consistent with the supernova observations \citep{sma09} and explosion simulations as discussed.
This means that $\sim6$ Myrs after the onset of star formation, stellar systems (globular clusters or super-early galaxies) experience the phase of less feedback and less oxygen enrichment than later times, depending on the hypernova fraction \citep[see also][]{ren25}.

There are nucleosynthesis yields with 2D jet-like explosions for hypernovae \citep{mae03,tom09}, but the results highly depend on the jet angle, which is a parameter. It is shown that multi-dimensional effects are particularly important for some elements, e.g., Sc, V, Ti, and Co. The `K15' GCE model is plotted in \citet{sne16},
applying constant factors, $+1.0$, 0.45, 0.3, 0.2, and 0.2 dex for [(Sc, Ti, V, Co, and $^{64}$Zn)/Fe] yields from \citet{kob11agb}, respectively, which take the 2D jet effects of HNe into account.

\begin{figure}[t]
\begin{center}
\includegraphics[width=0.7\textwidth]{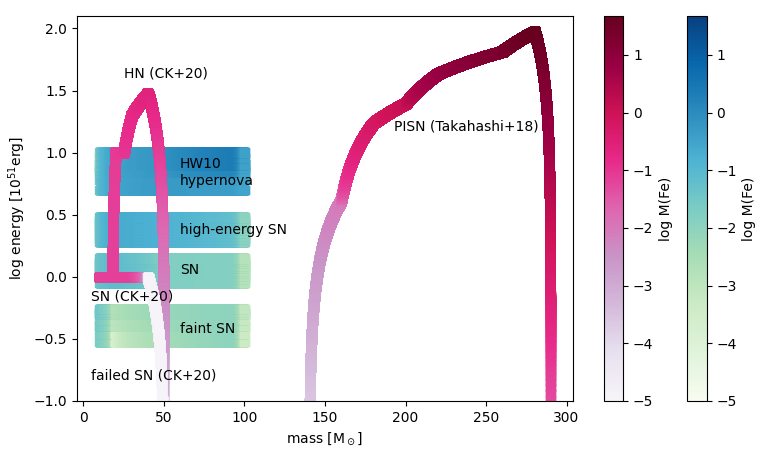}
\caption{\label{fig:snME}
Explosion energy as a function of initial mass, colour-coded with the ejected Fe mass
for the yields from \citet[blue to green]{heg10} and from K20 (red to purple) for $Z=0$. 
The observed mass--energy relation is assumed in K20.
Our {\it faint} supernovae follow the same relation but varying mixing-fallback to have $M({\rm Fe})\sim10^{-5}$--$10^{-2}M_\odot$ \citep{ish18}, and {\it failed} supernovae have $M({\rm Fe})=0$.
PISN yields are taken from \citet[no rotation, no magnetic fields]{tak18}.
}
\end{center}
\end{figure}

Figure \ref{fig:snME} should explain the key difference among 1D explosive nucleosynthesis calculations.
For core-collapse supernovae, explosion energy is a input parameter, and is given as a function of initial mass.
The ejected Fe mass (colour-coded) is not purely an output from the theoretical calculations as it is basically determined from the assumptions of mixing and/or fallback.

\citet{woo95} provided a yield table as a function of mass and metallicity, which was tested with a GCE model by \citet{tim95}, and has been widely used in chemical evolution studies. It would be useful to note that stellar mass loss is not included in their presupernova evolution calculations. Another problem is that their core-collapse supernova models tend to produce more Fe than observed because of the relatively deep mass cut, which leads to [$\alpha$/Fe]$\sim0$ in the ejecta. Therefore, the theoretical Fe yields were divided by a factor of 2 \citep{tim95,rom10}, which is obviously unphysical. This artificial reduction could mimic a shallower mass cut, but in that case the yields of other iron peak elements formed in the same layer as Fe should also be reduced. For the first problem, \citet{por98} obtained C+O core masses from stellar evolution models with mass loss and adopted to the \citet{woo95} supernova yields, which is not physically accurate either. It is also known that Mg production in the $40M_\odot$ model with $E_{51}=1$ is unreasonably small compared with that in other models. This Mg underproduction problem also exists in \citet{por98}.
\citet{heg10} provided a parameter study for 10--100$M_\odot$ varying many parameters such as explosion energy and mixing, without rotation, but only for $Z=0$.
These yields are used in \cite{vanni23} for constraining the first stars.
It is important to note that there is no guarantee that all of these models are realized in our Universe.
\citet{grimmett18} expanded the energy grid further at $Z=0$.

\citet{lim18} also provided a yield table for 13--120$M_\odot$ at four different metallicities ($Z>0$) as a function of stellar rotation. 
Explosive nucleosynthesis is calculated by imparting instantaneously an initial velocity $v_0$ to a mass coordinate in the iron core \citep{chi13}.
Mixing-fallback is also included in some sets, but with a different way from ours.
Their recommended set `R' yields are included in \citet{kob22iau}.
These yields do not reproduce the observed elemental abundances in the solar neighbourhood, namely iron-peak elements, which is probably because they do not include hypernovae.
Compared with observations, C is over-produced and Mg is under-produced, which might be due to some nuclear reaction rates.
\citet{roberti24b} added super-solar metallicity models, and \citet{roberti24a} added zero and extremely-low metallicity models but only for 15 and $25M_\odot$.

Stellar rotation induces mixing of C into the H-burning shell, producing a large amount of 
{\it primary} nitrogen, which is mixed back into the He-burning shell \citep{mey02,hir07}.
\citet{chi06} showed that rotation is necessary to explain the observed N/O--O/H relations with a GCE model, and the same result was shown in Fig.13 of \citet{kob11agb}.
However, using more self-consistent cosmological simulations, \citet{vin18no} reproduced the observed relation not with rotation but with inhomogeneous enrichment from AGB stars (\S \ref{sec:cosmo}).

For high initial rotational velocities at low metallicity (``spin stars''), this process results in the production of s-process elements, even at low metallicities \citep{fri16,lim18,cho18}; here the neutrons are mostly provided by the $^{22}$Ne($\alpha$,n)$^{25}$Mg reaction.
\citet{pra18} showed a GCE model assuming a metallicity-dependent function of the rotational velocities, and concluded that because of the contribution from rotating massive stars, a further light element primary process (LEPP) is not necessary to explain the elemental abundances with $A<100$.
However, in K20 GCE model, a significant amount of light neutron capture elements (Sr, Y, Zr) are produced by AGB stars and ECSNe, and thus adding stellar rotation results in overproduction of these elements \citep{kob22iau}.

There are also pre-supernova yields in the literature that do not include explosive nucleosynthesis. These are not useful for GCE since during explosions iron-peak elements are produced and $\alpha$ element yields are also largely modified, although they can be used for studying isotopic ratios of light elements.

\subsection{Low-mass stars}
\label{sec:agb}

Stars with initial masses between about 0.8--8$M_\odot$ (depending on
metallicity) pass through the thermally-pulsing AGB phase. The
He-burning shell is thermally unstable and can drive mixing between
the nuclearly processed core and envelope. This mixing is known as 
the third dredge-up (TDU), and is responsible for enriching
the surface in $^{12}$C and other products of He-burning,
as well as elements heavier than Fe produced by the {\em slow} neutron
capture process \citep{busso99,her05}.
Importantly, the TDU can result in the formation of a C-rich envelope,
where the C/O ratio in the surface layers exceeds unity.  In AGB stars
with initial masses $\gtsim 4M_\odot$, the base of the convective 
envelope becomes hot enough to sustain proton-capture nucleosynthesis 
(hot bottom burning, HBB).  HBB can change the surface 
composition because the entire envelope is exposed to the hot 
burning region a few thousand times per interpulse period. 
The CNO cycles operate to convert the freshly synthesized  $^{12}$C
into {\em primary} $^{14}$N, and the NeNa and MgAl chains may also
operate to produce $^{23}$Na and Al; this is one of the sites for the observed O-Na anti-correlation in globular clusters \citep{kraft97}).
Overall a large fraction of light elements such as C, N and F are
produced by AGB stars, while the contribution toward heavier elements 
(from Na to Fe) is negligible, except perhaps for specific isotopes 
(e.g., $^{22}$Ne, $^{25,26}$Mg), in GCE \citep[see][for the details]{kob11agb}.

At the deepest extent of each TDU, it is assumed that the bottom of the H-rich convective envelope penetrates into the $^{12}$C-rich intershell layer resulting in a partial mixing zone leading to the formation of a $^{13}$C pocket via the $^{12}$C(p,$\gamma$)$^{13}$N($\beta^+$)$^{13}$C reaction chain. While many physical processes have been proposed, there is still not full agreement on which process(es) drives the mixing. The inclusion of $^{13}$C pockets in theoretical calculations of AGB stars is still one of the most significant uncertainties affecting predictions of the s-process and in particular the absolute values of the yields \citep[][and references therein]{kar16,buntan17}. Other major uncertainties come from the rates of the neutron source reactions $^{13}$C($\alpha$,n)$^{16}$O and $^{22}$Ne($\alpha$,n)$^{25}$Mg \citep{bisterzo15} and the neutron-capture cross sections of some key isotopes \citep{ces18}.

The yields for AGB stars were originally calculated in \citet{kar10} and \citet{kar16}, but a new set with the s-process is used in K20 with optimising the mass of the partial mixing zone.
In K20, the narrow mass range of super-AGB stars is also filled with the yields from \citet{doh14a}; stars at the massive end are likely to become ECSNe.
At the low-mass end, off-centre ignition of C flame moves inward but does not reach the centre, which remains a hybrid C+O+Ne WD. This might become a sub-class of SNe Ia called SN Iax (\S \ref{sec:ia}).
There are other yields in the literature, see \citet{rom10} and \citet{rom19} for comparison with GCE.

\subsection{Type Ia Supernovae}
\label{sec:ia}

\begin{figure}[t]
\begin{center}
\includegraphics[width=0.6\textwidth]{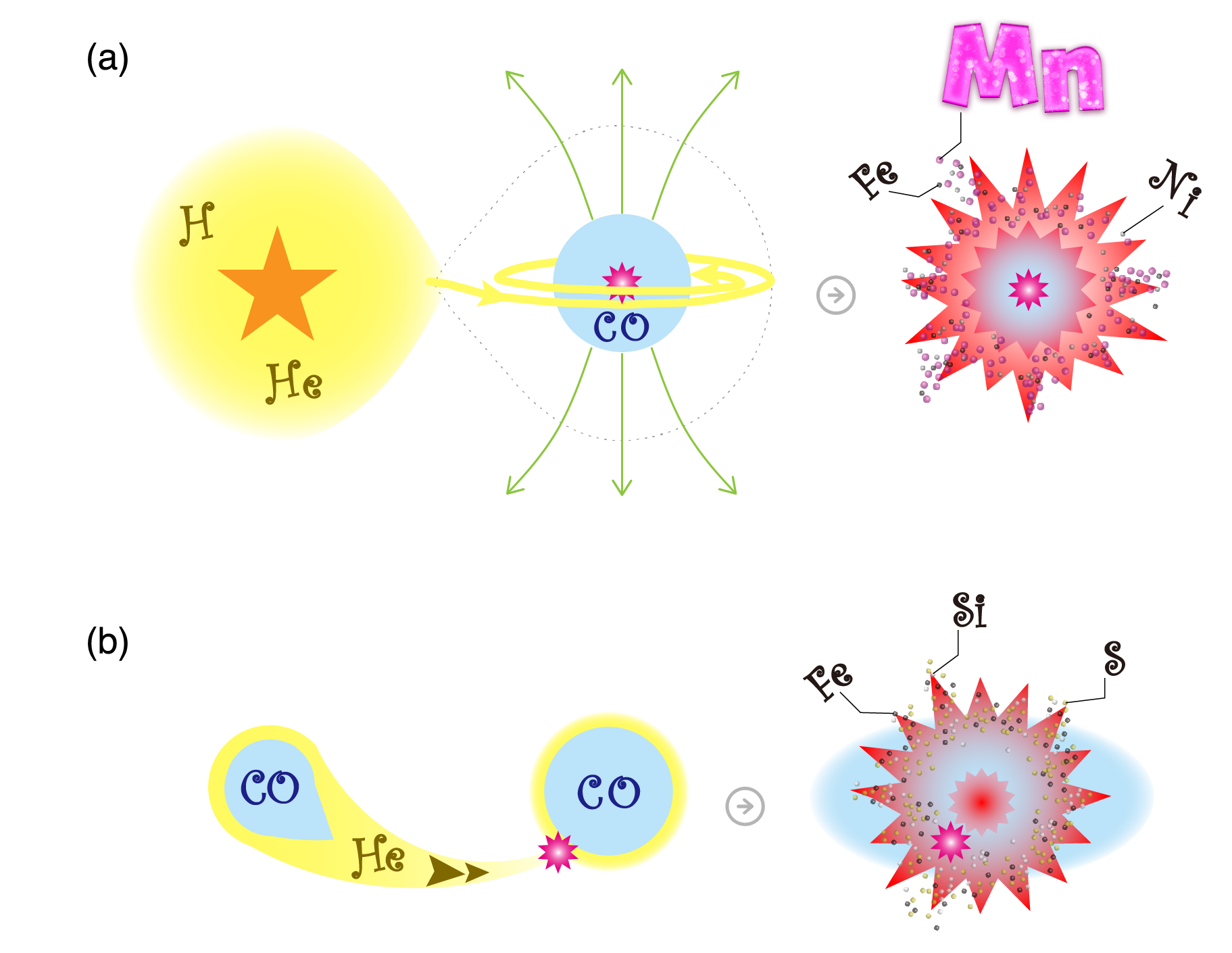}
\caption{\label{fig:ia}
Schematic view of SN Ia progenitors and their explosions. 
(a) Near-Ch mass SN Ia in a single degenerate system, which sufficiently produces Mn.
(b) Sub-Ch mass SN Ia in a double degenerate system.
[Credit: C. Kobayashi \& IPMU 2020]
}
\end{center}
\end{figure}

Although Type Ia Supernovae (SNe Ia) have been used as a standard candle to measure the expansion of the Universe, the progenitor of SNe Ia is still a matter of big debate \citep[e.g.][for a review]{ruiter24}. It is a combined problem of the progenitor systems and the explosion mechanism \citep{kob20ia}.
Among several scenarios proposed,
Figure \ref{fig:ia} shows (a) near-Chandrasekhar (Ch)-mass explosions of carbon-oxygen (C+O) WDs that are expected in single degenerate systems,
and (b) sub-Ch-mass explosions that can happen in both single and double degenerate systems.

For the nucleosynthesis yields, a deflagration model W7 of a Ch-mass WD \citep{nom84,iwa99} has been the most favoured 1D model for reproducing the observed spectra of SNe Ia \citep{hof96,nug97}.
In recent works, 3D simulations of a delayed detonation (DDT) in a Ch-mass WD and of a violent merger of two WDs \citep{roepke12}, and 2D simulations of a double detonation in a sub-Ch-mass WD \citep{kro10} can also give a reasonable match with observations.
The advantage of the W7 model is that it also reproduces the GCE in the solar neighborhood, namely, the observed increase of Mn/Fe with metallicity as well as the decrease of $\alpha$ elements (O, Mg, Si, S, and Ca).
However, the Ni/Fe ratio is too high at [Fe/H] $\gtsim -1$ \citep{kob06}.
An updated GCE model with the DDT yields from \citet{sei13} was shown in \citet{sne16}, which indeed gives Ni/Fe ratios closer to the observational data.
In contrast to these Ch-mass models, sub-Ch-mass models, which have been re-considered for SNe Ia with a number of other observational results such as supernova rates \citep[e.g.,][]{mao14} and the lack of donors in supernovae remnants \citep{ker09}, do not match the GCE in the solar neighborhood.
However, the Mn production from sub-Ch-mass models is too small to explain the observations in the solar neighborhood \citep{sei13}.
SNe Iax could compensate this with their large Mn production.
The contribution of SN Iax is included in GCE by \citep{kob15,kob20sr}, is found to be negligible in the solar neighbourhood, but can be important for dwarf spheroidal galaxies \citep{kob15,ces17}.

\citet{kob20ia} newly calculated the nucleosynthesis yields of near-Ch and sub-Ch mass models, using the same 2D code as in \citet{leu18} and \citet{leu20} but with more realistic, solar-scaled initial composition. The initial composition gives significantly different (Ni, Mn)/Fe ratios, compared with the classical W7 model \citep{nom97ia} or more recent delayed detonation model \citep{sei13}.
When constraining the progenitors of SNe Ia from the observed Mn abundances in the solar neighborhood, it is important to use the latest yields of SNe Ia.
See \citet{blondin22} for a complete comparison including many other yields.

By comparing observed stellar abundances in the solar neighbourhood, \citet{kob20ia} found that the contribution from Ch-mass explosions should be as high as 75\%.
This is higher than in \citet{sei13}, which had higher Mn yields from Ch-mass models probably due to artificial initial compositions of WDs.
The relative contribution also depends on the core-collapse-supernova yields, which determine (Mn,Ni)/Fe at low metallicities in the model, as well as non-local thermodynamic equilibrium (NLTE) corrections for Mn and Ni abundances in observations.
Moreover, the relative contribution may depend on the environments; \citet{kir19} and \citet{kob20ia} found that sub-Ch mass SNIa should be dominant in satellite dwarf spheroidal (dSph) galaxies of the Milky Way.

\subsection{The r-process sites}

Elements beyond iron contain more neutrons than protons in their nuclei, and are produced by neutron-capture processes in the neutron-rich matter.
If the neutron density $N_{\rm n} \sim 10^7$ cm$^3$, the neutron capture occurs on a longer timescale than $\beta$ decays following the path just below the $\beta$ stability valley in the nuclear chart. 
This s-process can happen in AGB stars (\S \ref{sec:agb}), and has the abundance pattern with three peaks around $A=$88 (Sr), 138 (Ba), 208 (Pb) corresponding to the magic numbers of nuclear physics.
If $N_{\rm n} > 10^{20}$ cm$^3$, the neutron capture occur much faster, far from the $\beta$ stability valley \citep{bbfh57}. The abundance pattern has peaks around $A=$80 (Se), 130(Xe), 195(Pt).
The r-process can also form the heaviest stable elements Th and U, and terminates at $A\sim270$ due to fission.
Note that a weak s-process can happen in rotating massive stars (\S \ref{sec:sn}), and there is also an intermediate process (i-process).

The astrophysical sites where the r-process occurs have been debated for half-a-century, with the main contenders being core-collapse supernovae and NS mergers \citep[e.g.,][]{cam73}.
The existence of a NS merger has been observationally confirmed by the gravitational wave (GW) source GW170817, also observed as an optical transient and as a short GRB.
However, Galactic chemical evolution models have been challenging NS mergers as the main site \citep{mat14,ces14,weh15,cot19,kob20sr}. Also in hydrodynamical simulations, an r-process associated with core-collapse supernovae is required \citep{hay19,van20}.
This is further supported by the discovery of Yong star at [Fe/H] $=-3.5$ \citep[\S \ref{sec:first}]{yon21a}, which indicated the supernova was relatively bright (magneto-rotational hypernovae).
Note that, collapsars \citep{sie19} can also give rapid enrichment as GCE models require, but if there is no supernova explosion, no Fe production, and thus the scatter of r-process elements relative to Fe becomes too large. It is also debated if the accretion disk around the BH can be neutron-rich or not \citep{just22}.

In K20, the r-process yields are taken from the following references: $8.8M_\odot$ model in \citet[][2D]{wan13ec} for electron capture supernovae, $1.2-2.4M_\odot$ models in \citet{wan13nu} for $\nu$-driven winds, $1.3M_\odot$+$1.3M_\odot$ model in \citet[][3D-GR]{wan14} for neutron star mergers, and $25M_\odot$ ``b11tw1.00'' model in \citet[][axisymmetric MHD]{nis15} for magneto-rotational supernovae.
These yields are uncertain depending on 1) initial conditions (e.g., $M$, $Z$, rotation, magnetic fields), 2) the quality of the base hydrodynamical simulations (3 dimensional general relativity or not), 3) neutrino physics, and 4) nuclear physics (e.g., reaction rates, fission modelling).

Also note that \citet{wan14}'s yields includes dynamical ejecta only, and the contribution from $\nu$-driven winds from BH torus may be important for relatively light n-capture elements such as Eu. Moreover, the yield dependence on the mass of the compact objects seems significant. Although there are no self-consistent nucleosynthesis yields from the simulations that can follow both dynamical ejecta and long-term outflows, \citet{kob22} showed the impact combining the yields from \citet{jus15}.

\subsection{Very massive stars}
\label{sec:vms}

\begin{figure}[t]
\begin{center}
\includegraphics[width=0.7\textwidth]{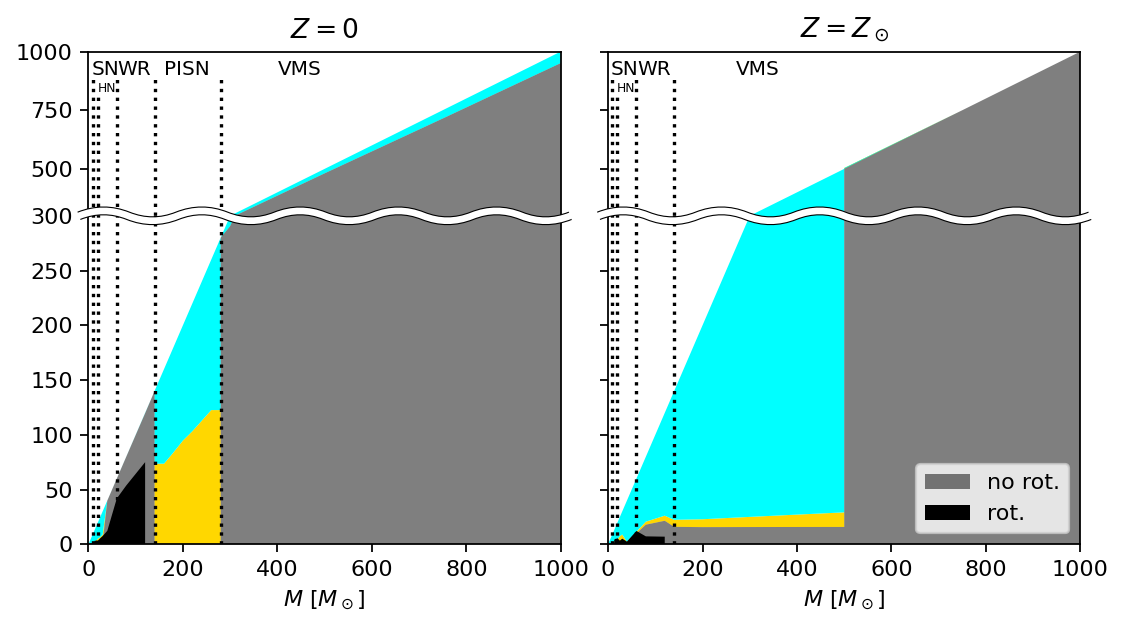}
\caption{\label{fig:vms}
The remnant and ejection masses (in $M_\odot$) as a function of initial mass of progenitor stars for the initial metallicity $Z=0$ (left) and $Z=Z_\odot$ (right).
Gold, cyan, gray shaded area are for metals, and H+He, and the remnant, respectively for non-rotating stars. The black areas show the remnant mass for rotating stars.
The yield data are taken from \citet{tak18,volpato23,higgins23} and K20, and are included in GCE models in \citet{kob24}.
}
\end{center}
\end{figure}

Very massive stars (VMSs, $>$100$M_\odot$) have been solidly identified in the Tarantula Nebula of the Large Magellanic Cloud \citep{crowther10,schneider18,bestenlehner20,brands22} and are included in GCE models of \citet{kob24} with the yields summarized in Figure \ref{fig:vms}.

Various stellar evolution models \citep[e.g.,][]{szecsi22}, and some nucleosynthesis yields \citep{yusof13,martinet22,volpato23} exist, although the results significantly depend on the input physics, namely stellar mass loss.
With \citet{vink11}'s enhanced winds, solar metallicity 100--500$M_\odot$ VMSs loose mass to become like $\sim30M_\odot$ stars \citep{higgins23}.
N production is seen as a result of the CNO cycle during core H-burning, as for Na, Al, $^{26}$Al.
These yields do not include explosive nucleosynthesis (i.e., stellar winds only).

The case with $Z=0$ has been intensively studied as the first stars were expected to be massive, with masses $\approx100M_\odot$ \citep[e.g.,][]{abel02,bromm04}.
However, the initial stellar mass depends on complex physical processes, such as gas fragmentation, ionization, accretion, and feedback from newborn stars. Once these processes are included in modern numerical simulations, it seems possible to form lower-mass stars \citep{greif11,hirano14}, and even binaries (\citealt{stacy13}; see also \citealt{hartwig23} for observational signatures).

If the masses of the first stars are $\sim$160--280$M_\odot$, they undergo thermonuclear explosions triggered by pair-creation instability \citep{barkat67,heg02,nom13,tak18} leaving no remnant. 
Pair-instability supernovae (PISNe) have a very distinct nucleosynthetic pattern. Although considerable effort has been made to detect such characteristic pattern, no observational signature for the existence of PISNe has yet been convincingly found, neither in EMP stars \citep{cay04,aguado23} nor in DLAs \citep[][\S \ref{sec:first}]{kob11dla,saccardi23}.

Stars above $\sim$90$M_\odot$ undergo nuclear instabilities and associated pulsations at various nuclear burning stages \citep{heg02}.
The evolution and nucleosynthesis of pulsating pair-instability supernovae (PPISN) are uncertain, but they are unlikely to explode and will leave a $\sim$100$M_\odot$ BH.

\subsection{Super-massive stars}

Super-massive stars (SMSs) were originally defined as those that cause GR instability before H burning \citep{ful86,woods19}.
They might form from gas clouds if molecular hydrogen is dissociated by Lyman-Werner
radiation, or by collisions in a dense star cluster \citep{rees84}, so could be metal-enhanced.
The structure and evolution also depend on their initial masses and their formation path, and thus their nucleosynthesis yields (\citealt[][$Z=5\times10^{-4}$]{denissenkov14};
\citealt[][$Z>0$]{nagele23}; \citealt[][$Z=0$]{nandal24}) are very uncertain.
These models have to assume mass loss as no observations can constrain it, as well as dilution, which determines the metallicities of the observed objects that could be enriched by SMSs: e.g., globular clusters, GN-z11 (\S \ref{sec:firstgal}).

$\sim10^4M_\odot$ stars have incomplete core H burning. These stars have been proposed as an explanation for the abundance anomaly \citep{denissenkov14}, and the so-called O--Na anti-correlation \citep{kraft97} often seen in globular clusters of the Milky Way.
$\sim10^5M_\odot$ stars are fully convective, while accreting ones are mostly radiative \citep{woods19}.
In a very narrow mass range these stars may explode as general relativistic supernovae \citep[GRSN;][]{chen14}.

\section{Galactic chemical evolution}
\label{sec:gce}

Galactic chemical evolution (GCE) has been calculated analytically and numerically since the 70's \citep[e.g.,][hereafter K00]{tinsley80,pra93,tim95,pagel97,chi97,matteucci01,kob00} basically by integrating the following equation:
\begin{equation}\label{eq:gce}
\frac{d(Z_if_{\rm g})}{dt}=E_{\rm SW}+E_{\rm SNcc}+E_{\rm SNIa}+E_{\rm NSM}-Z_i\psi+Z_{i,{\rm inflow}}R_{\rm inflow}-Z_iR_{\rm outflow}
\end{equation}
where the mass fraction of each element $i$ in gas-phase ($f_{\rm g}$ denotes the gas fraction, or the gas mass in the system considered with a unit mass) increases via element ejections from stellar winds ($E_{\rm SW}$), core-collapse supernovae ($E_{\rm SNcc}$), Type Ia supernovae ($E_{\rm SNIa}$), and neutron star mergers ($E_{\rm NSM}$). It also decreases by star formation (with a rate $\psi$), and can be modified by inflow (with a rate $R_{\rm inflow}$) and outflow (with a rate $R_{\rm outflow}$) of gas in/from the system considered.
It is assumed that the chemical composition of gas is instantaneously well mixed in the system (called an one-zone model), but the instantaneous recycling approximation is not adopted nowadays \citep{matteucci21}.
The model with $R_{\rm inflow}=R_{\rm outflow}=0$ is called a closed-box model, but is not realistic in any observed galaxies.
The initial conditions are $f_{{\rm g},0}=1$ (a closed system) or $f_{{\rm g},0}=0$ (an open system) with the chemical composition ($Z_{i,0}$) from the Big Bang nucleosynthesis ($Z_{i,{\rm BBN}}$). External enrichment is often neglected, assuming $Z_{i,{\rm inflow}}=Z_{i,{\rm BBN}}$.
In Eq.(\ref{eq:gce}) the first two terms depend only on nucleosynthesis yields, while the third and fourth terms also depend on modelling of the progenitor binary systems, which is uncertain. The last three terms are galactic terms, and should be determined from galactic dynamics, but are assumed with analytic formulae in GCE models.
The stars formed at a given time $t$ have the initial metallicity, $Z(t)$, which is equal to the metallicity of the ISM from which the stars form\footnote{Later, integrated metallicity of stellar populations in a galaxy will be described as $Z_*$.}.

\subsection{The metal ejection terms in GCE}

Nucleosynthesis yields ($p_{z_im,{\rm X}}$) are integrated over stellar lifetimes of single stars, or delay time distributions for binaries, depending on $Z$ (Eqs.\,\ref{eq:e_sw}--\ref{eq:e_nsm}). It is extremely important to take account of this metallicity dependence, though it is often ignored in hydrodynamical simulations. Because of this, it is also not possible to solve chemical evolution as a post-process. Chemical enrichment including all elements must be followed {\it on-the-fly}.

The ejection terms are given by integrating the contributions with various initial mass ($m$) and metallicity ($Z$) of progenitor stars. Hence, the first assumption is the initial mass function (IMF), $\phi$.
The IMF is often assumed to have time/metallicity-invariant mass spectrum normalized to unity at $m_\ell \leq m \leq m_u$ as
\begin{equation}
\phi(m) \propto m^{-x} ,
\end{equation}
\begin{equation}
\int_{m_\ell}^{m_u} m^{-x} dm= \frac{1}{1-x} (m_u^{1-x}-m_\ell^{1-x}) = 1 .
\end{equation}
\citet{sal55} found the slope of the power-law, $x=1.35$, from observation of nearby stars.
The slope depends on the exact definition of the IMF \citep{tinsley80}, and the IMF per number is $\alpha=x+1=2.35$.
\citet{kro93} updated this result finding three slopes at different mass ranges, which is summarised as $x=-0.7, 0.3$, and $1.3$ for $m/M_\odot \ltsim 0.08, 0.08 \ltsim m/M_\odot \ltsim0.5$, $0.05\ltsim m/M_\odot \ltsim 150$, respectively \citep[Eq.2 of][]{kro01,kro08}. However, the author suggested that the slope of the massive end should be $x=1.7$ in the solar neighborhood, which has been used in some GCE works \citep[e.g.,][]{rom10}.
In most of results in this paper the Kroupa IMF (with the massive-end slope $x=1.3$) is adopted for a mass range from $m_\ell=0.01 M_\odot$ to $m_u=120 M_\odot$, while \citet{cha03}'s IMF is often used in hydrodynamical simulations.
Note that there are a few claims for an IMF variation from observations \citep[e.g.,][]{vandokkum10,gunawardhana11,cappellari13}.

Then the ejection terms for single stars can be calculated as follows\footnote{The metals in Eq.(\ref{eq:e_sw}) and Eq.(\ref{eq:e_cc}) are called `unprocessed' and `processed' metals, respectively. Our supernova yield table contains `processed' metals only, the AGB yield table contains net yields, so that the mass loss with star particle's chemical composition should also be included with Eq.(\ref{eq:e_sw}).}.
\begin{equation}
E_{\rm SW}=\int_{m_t}^{m_u}\,(1-w_m-p_{z_im,{\rm II}})\,Z_i(t-\tau_m)\,
\psi(t-\tau_m)\,\phi(m)~dm ,
\label{eq:e_sw}
\end{equation}
\begin{equation}
E_{\rm SNcc}=\int_{m_t}^{m_u}\,p_{z_im,{\rm II}}\,
\psi(t-\tau_m)\,\phi(m)~dm .
\label{eq:e_cc}
\end{equation}
The lower mass limit for integrals is the turn-off mass $m_t$ at $t$, which
is the mass of the star with the main-sequence lifetime $\tau_m=t$.
$\tau_m$ is a lifetime of a star with $m$, and also depends on $Z$ of the star.
$w_m$ is the remnant mass fraction, 
which is the mass fraction of a WD (for stars with initial masses of $m\ltsim8M_\odot$), a NS (for $\sim 8-20M_\odot$) or a black hole (BH) (for $\gtsim20M_\odot$). 
$p_{z_im,{\rm II}}$ denotes the nucleosynthesis yields of core-collapse supernovae, given as a function of $m$, $Z$, and explosion energy.
The star formation rate $\phi(t)$ at a time $t$ from the formation epoch is given in the next subsection.

The additional ejection terms from binary systems are more complicated. Suppose that the event rates are given, the enrichment can be calculated as
\begin{equation}
E_{\rm Ia}=m_{\rm CO}\,p_{z_im,{\rm Ia}}\,{\cal R}_{t,{\rm Ia}} ,
\label{eq:e_ia}
\end{equation}
\begin{equation}
E_{\rm NSM}=m_{\rm NSM\,ejecta}\,p_{z_im,{\rm NSM}}\,{\cal R}_{t,{\rm NSM}} .
\label{eq:e_nsm}
\end{equation}
All of the matter in the WD ($m_{\rm CO}$) is ejected from SNe Ia (except for a sub-class called SNe Iax, see \citealt{kob15}); for Ch-mass SNe Ia, $m_{\rm CO}=1.38M_\odot$.
On the other hand, only a small fraction of matter is ejected from a NSM ($m_{\rm NSM\,ejecta}\sim0.01M_\odot$), depending on the mass ratios, the equation of state of the NSs, and the spin of BHs \citep{kob22}.
These are provided together with the nucleosynthesis yields ($p_{z_im,{\rm Ia}}$ and $p_{z_im,{\rm NSM}}$), depending on the progenitor system.

${\cal R}_{t,{\rm SNIa}}$ and ${\cal R}_{t,{\rm NSM}}$ are the rates of SNe Ia and NSMs per unit time per unit stellar mass formed in a population of stars with a coeval chemical composition and age (simple stellar population, SSP), and are also called the delay-time distribution (DTD).
For the DTDs, simple analytic formula were also proposed \citep[][and references therein]{matteucci21}. However, the functions are significantly different from what are obtained from binary population synthesis (BPS), ignore metallicity dependences during binary evolution, and do not includes the effects of supernova kicks for NSMs. \citet{ded04} was the first work that combined BPS to GCE.

For SNe Ia from the single-degenerate systems, \cite{kob98} proposed another analytic formula based on binary calculation:
\begin{equation}
{\cal R}_{t,{\rm Ia}}=b~
\int_{\max[m_{{\rm p},\ell}(Z),\,m_t]}^{m_{{\rm p},u}(Z)}\,
\frac{1}{m}\,\phi(m)~dm~
\int_{\max[m_{{\rm d},\ell}(Z),\,m_t]}^{m_{{\rm d},u}(Z)}\,
\frac{1}{m}\,\psi(t-\tau_m)\,\phi_{\rm d}(m)~dm .
\label{eq:r_ia}
\end{equation}
The first integral is for the primary star, and the mass range is for the stars that can produce $\sim 1M_\odot$ of C+O WDs in binaries, which is set to be $\sim 3-8M_\odot$.
The second integral is for the secondary star, and the mass range depends on the optical thick winds from the WD, which is about $\sim1M_\odot$ and $\sim3M_\odot$ for the red-giant+WD systems and main-sequence+WD systems, respectively, depending on $Z$.

The metallicity dependence on the SN Ia DTD is very important to reproduce the observed [$\alpha$/Fe]--[Fe/H] relation in the solar neighbourhood \citep[see][for the detailed parameters and the resultant DTDs]{kob09}.
These parameters are for Ch-mass explosions. Eq.(\ref{eq:r_ia}) can also be used for SNe Iax and sub-Ch mass explosion triggered by slow H accretion \citep[see][for the detailed parameters]{kob15}, but not for the double-degenerate systems (which are likely to cause sub-Ch mass explosions).
DTDs of these sub-classes of SNe Ia have been predicted by various BPS models, but none of these models can reproduce the observations because the total rate is too low and/or the typical timescale is too short \citep[see][for more details]{kob22}.

For NSMs, K20 used a metallicity-dependent DTD from a BPS \citep{men14,men16}.
\citet{kob22} used various BPS models and also provided new analytic formulae that can reproduce the observed [Eu/(Fe,O)]--[(Fe,O)/H] relation only with NSMs, without the r-process associated with core-collapse supernovae.
Currently, there is no BPS model that can explain the observation only with NSMs because the rate is too low and/or the timescale is too long \citep[see][for more details]{kob22}.
Other binary systems such as novae can also contribute GCE for some elements or isotopes \citep[e.g.,][]{rom19} but the DTDs are very uncertain \cite[see][for the nova DTDs from BPS]{kem22} and the yield tables are not available.

\begin{figure}[t]
\begin{center}
\includegraphics[width=0.7\textwidth]{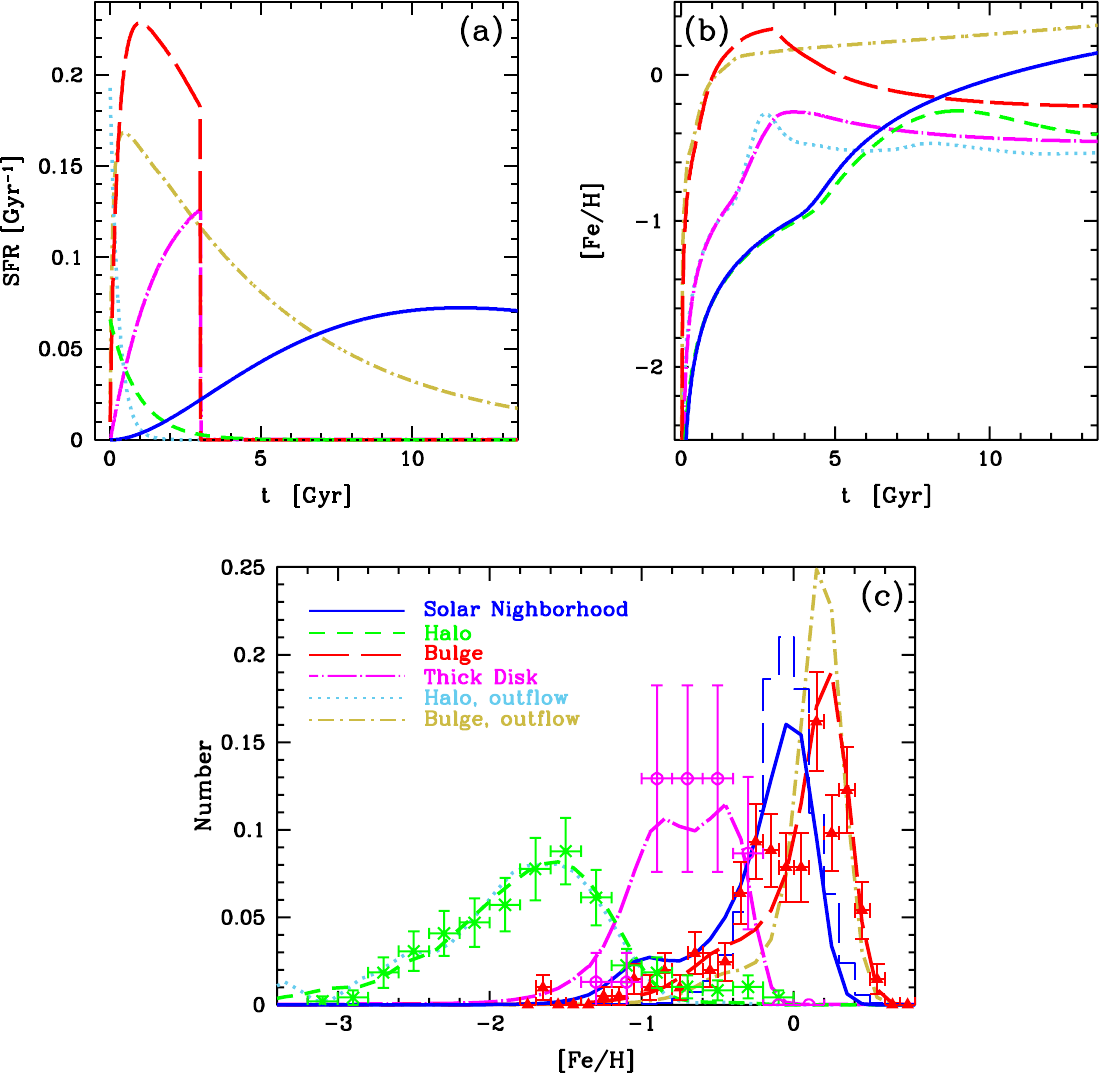}
\caption{\label{fig:mdf}
Star formation histories (panel a), age-metallicity relations (panel b), and metallicity distribution functions (panel c) for the 
solar neighborhood (blue solid lines),
halo (green short-dashed lines),
halo with stronger outflow (light-blue dotted lines),
bulge (red long-dashed lines),
bulge with outflow (olive dot-short-dashed lines)
and thick disk (magenta dot-long-dashed lines).
Figure is taken from \citet{kob20sr}, see the reference for the observational data sources and the model details.
}
\end{center}
\end{figure}

\begin{figure}[t]
\begin{center}
\includegraphics[width=0.7\textwidth]{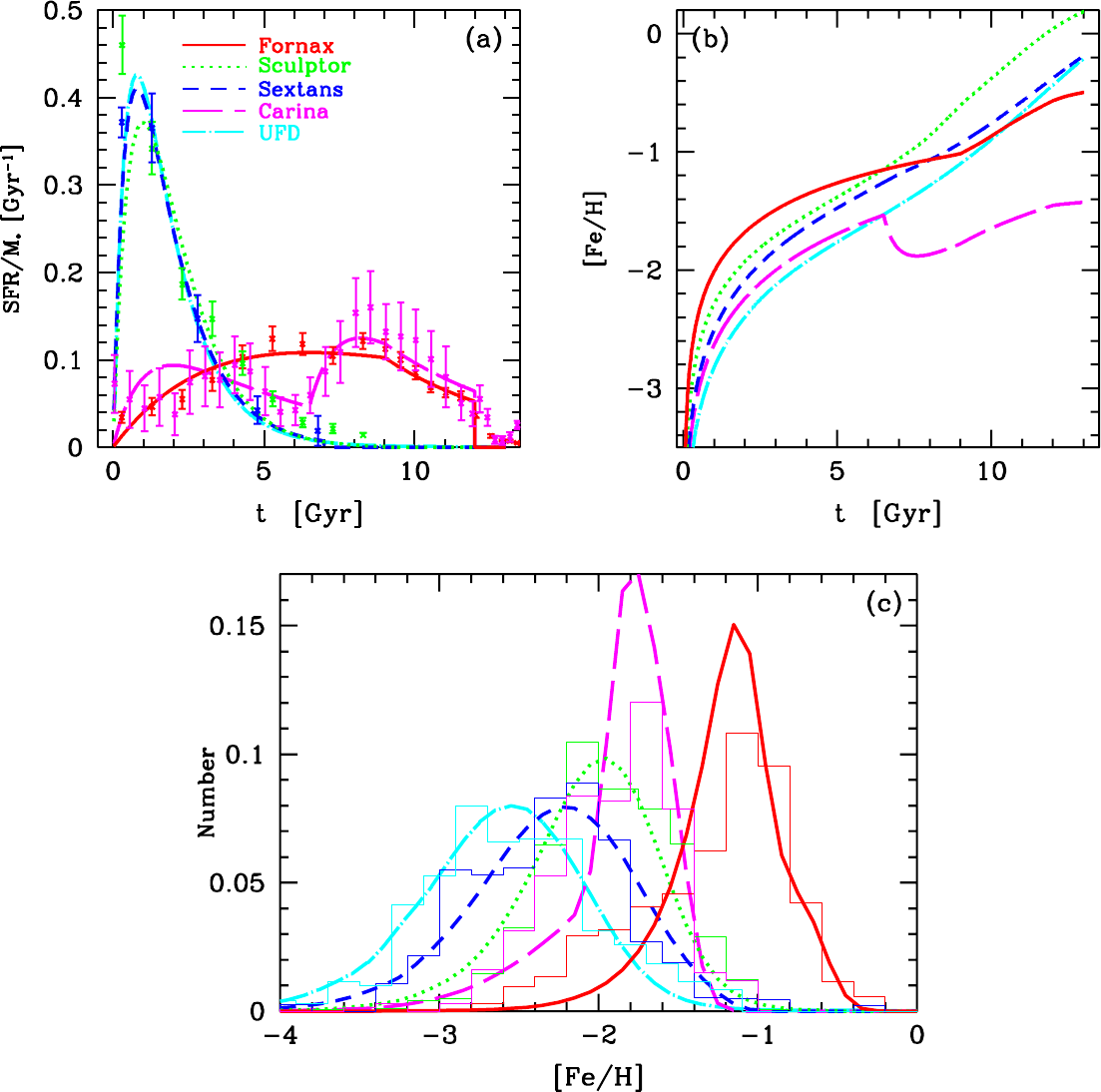}
\caption{\label{fig:mdf2}
The same as Fig.\,\ref{fig:mdf} but for dwarf spheroidal galaxies, Fornax (red solid lines), Sculptor (green long-dashed lines), Sextans (blue short-dashed lines), and Carina (magenta dotted lines), and ultra-faint dwarf (UFD) galaxies (cyan dot-dashed lines).
Figure is updated from \citet{kob20ia}, see the reference for the observational data sources and the model details.
The observational data for the UFD is provided by A. Ji (priv. comm.).
}
\end{center}
\end{figure}

\subsection{The galactic terms in GCE}

From the observed Kennicutt–Schmidt law \citep{ken12}, the star formation rate is usually modelled as
\begin{equation}
\psi\!=\!\frac{1}{\tau_{\rm s}}f_{\rm g}^{\,n}
\end{equation}
but often $n=1$ is adopted, and thus it is proportional to the gas fraction.
${\tau_{\rm s}}$ denotes a timescale of star formation, and $1/{\tau_{\rm s}}$ gives the efficiency of star formation.
Note that star formation may be more complicated depending on turbulence and magnetic fields \citep[e.g.,][]{fed11}.

The inflow rate is often assumed to be exponential as
\begin{equation}
R_{\rm inflow}\!\!=\!\frac{1}{\tau_{\rm i}}\exp\frac{-t}{\tau_{\rm i}}
\end{equation}
but occasionally as 
\begin{equation}
R_{\rm inflow}\!\!=\!\frac{t}{\tau_{\rm i}^2}\exp\frac{-t}{\tau_{\rm i}}
\end{equation}
in \citet{pagel97} and in the solar neighbourhood model of K20.
${\tau_{\rm i}}$ denotes a timescale of inflow.
For the cosmological inflow, the chemical composition is primordial: $Z_{\rm inflow}=0$.
If the inflow is driven by galaxy mergers or `fountain' (matter ejected from the disk falls back on the disk), the infalling matter contains a non-negligible amount of metals: $Z_{\rm inflow}>0$.
For the radial flow, the gas cools and flows inward on the disk plane where star formation is ongoing, and thus the metallicity can be high: $Z_{\rm inflow}>>0$, which can steepen metallicity radial gradients of galaxies.
\citet[][\S \ref{sec:mw}]{vin20} found this radial flow has a high-$\alpha$ abundance while the fountain is relatively low-$\alpha$ due to SNe Ia.

The outflow rate is assumed to be proportional to the star formation rate as
\begin{equation}
R_{\rm outflow}\!\!=\!\frac{1}{\tau_{\rm o}}f_{\rm g}^{\,n}\propto\psi ,
\end{equation}
which is reasonable if the outflow is driven by supernova feedback. ${\tau_{\rm o}}$ denotes a timescale of outflow. Alternatively, star formation is quenched and is simply set to $\psi=0$ after a given epoch $t_{\rm w}$, which corresponds to a galactic wind driven by the feedback from active galactic nuclei (AGN).
\citet[][\S \ref{sec:cosmo}]{tay20} found that the metal-loss due to AGN-driven winds are important for massive galaxies.
For satellite dwarf galaxies, mass loss due to tidal or ram-pressure stripping may also be important.

In GCE modelling it is extremely important to show the metallicity distribution function (MDF) and compare with observations.
In K20, the timescales are determined to match the observed MDF of stars in each system modelled.
Figures \ref{fig:mdf} and \ref{fig:mdf2} show the assumed star formation history (SFH) and the resultant metallicity evolution and MDF.
The parameter sets that have very similar MDFs give almost identical tracks of elemental abundance ratios (Fig.\,A1 of \citealt{kob20ia}). This means that, for the given MDF, elemental abundance tracks do not depend so much on the SFH.
Therefore, if the MDF is known, GCE can be used to constrain nuclear astrophysics from the Galactic archaeology.
Without knowing the MDFs, SFHs are unconstrained, and elemental abundance tracks can vary depending on the SFH.
Provided that nuclear astrophysics is known, elemental abundance measurements can be used to constrain the SFH through GCE, which is the case for extra-galactic archaeology.

\subsection{The origin of elements}

\begin{figure*}[t]
\begin{center}
  \includegraphics[width=0.95\textwidth]{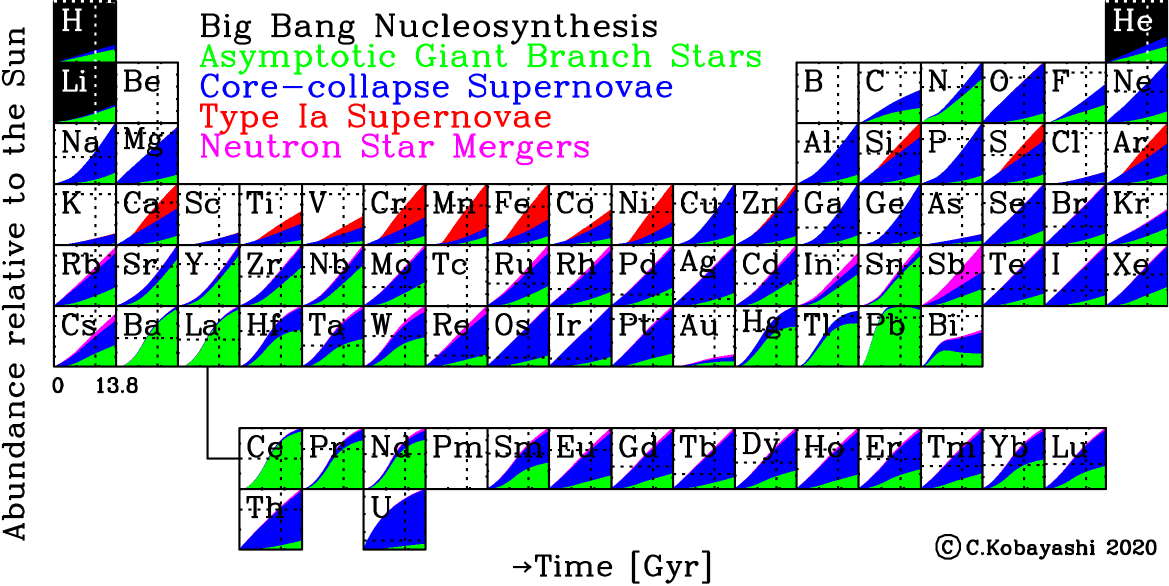}
\caption{The time evolution (in Gyr) of the origin of elements in the periodic table: Big Bang nucleosynthesis (black), AGB stars (green), core-collapse supernovae including SNe II, HNe, ECSNe, and MRSNe (blue), SNe Ia (red), and NSMs (magenta). The amounts returned via stellar mass loss are also included for AGB stars and core-collapse supernovae depending on the progenitor mass. The dotted lines indicate the observed solar values.
Figure is taken from \citet{kob20sr}.
}
\label{fig:origin}
\end{center}
\end{figure*}

Using the K20 GCE model for the solar neighbourhood, we summarize the origin of elements in the form of a periodic table. In each box of Figure \ref{fig:origin}, the contribution from each chemical enrichment source is plotted as a function of time: Big Bang nucleosynthesis (black), AGB stars (green), core-collapse supernovae including SNe II, HNe, ECSNe, and MRSNe (blue), SNe Ia (red), and NSMs (magenta).
It is important to note that the amounts returned via stellar mass loss are also included for AGB stars and core-collapse supernovae depending on the progenitor star mass (Eq.\ref{eq:e_sw}).
The x-axis of each box shows time from $t=0$ (Big Bang) to $13.8$ Gyrs, while the y-axis shows the linear abundance relative to the Sun, $X/X_\odot$.
The dotted lines indicate the observed solar values, i.e., $X/X_\odot=1$ and $4.6$ Gyr for the age of the Sun;
the solar abundances are taken from \citet{asp09}, except for $A_\odot({\rm O}) = 8.76,  A_\odot({\rm Th}) = 0.22$, and $A_\odot({\rm U}) = -0.02$ (\S 2.2 of K20 for the details).
The adopted star formation history is similar to the observed cosmic star formation rate history, and thus this figure can also be interpreted as the origin of elements in the universe, which can be summarized as follows:
\begin{itemize}
\item H and most of He are produced in Big Bang nucleosynthesis. As noted, the green and blue areas also include the amounts returned to the ISM via stellar mass loss in addition to He newly synthesized in stars.
After tiny production in Big Bang nucleosynthesis,
Be and B are supposed to be produced by cosmic rays \citep{pra93}, 
which are not included in the K20 model.
\item The Li model is very uncertain because the initial abundance and nucleosynthesis yields are uncertain. Li is supposed to be produced also by cosmic rays and novae, which are not included in the K20 model. The observed Li abundances show an increasing trend from very low metallicities to the solar metallicity, which could be explained by cosmic rays. Then the observation shows a decreasing trend from the solar metallicities to the super-solar metallicities, which might be caused by the reduction of the nova rate \citep{gri19}; this is also shown in theoretical calculation with binary population synthesis \citep{kem22}, where the nova rate becomes higher due to smaller stellar radii and higher remnant masses at low metallicities.
\item 49\% of C, 51\% of F, and 74\% of N are produced by AGB stars\footnote{In extra-galactic studies, the N production from AGB stars is referred as ``secondary'', which is confusing, and is a {\it primary} production from freshly synthesized $^{12}$C. For massive stars, the N yield depends on the metallicity of progenitor stars for {\it secondary} production, and can also be enhanced by stellar rotation for {\it primary} production \citep{kob11agb}.} (at $t=9.2$ Gyr).
For the elements from Ne to Ge, the newly synthesized amounts are very small for AGB stars, and the small green areas are mostly for mass loss.
\item $\alpha$ elements (O, Ne, Mg, Si, S, Ar, and Ca) are mainly produced by core-collapse supernovae, but 22\% of Si, 29\% of S, 34\% of Ar, and 39\% of Ca come from SNe Ia. 
These fractions would become higher with sub-Ch-mass SNe Ia \citep{kob20ia} instead of 100\% Ch-mass SNe Ia adopted in the K20 model.
\item A large fraction of Cr, Mn, Fe, and Ni are produced by SNe Ia. In classical works, most of Fe was thought to be produced by SNe Ia, but the fraction is only 60\% in the K20 model, and the rest is mainly produced by HNe. The inclusion of HNe is very important as it changes the cooling and star formation histories of the universe significantly \citep{kob07}.
Co, Cu, Zn, Ga, and Ge are largely produced by HNe.
In the K20 model, 50\% of stars at $\ge 20M_\odot$ are assumed to explode as hypernovae, and the rest of stars at $> 30M_\odot$ become failed supernovae.
\item Among neutron-capture elements, as predicted from nucleosynthesis yields, AGB stars are the main enrichment source for the s-process elements at the second (Ba) and third (Pb) peaks. 
\item 32\% of Sr, 22\% of Y, and 44\% of Zr can be produced from ECSNe, which are included in the blue areas, even with the adopted conservative mass ranges; we take the metallicity-dependent mass ranges from the theoretical calculation of super-AGB stars \citep{doh15}.
Combined with the contributions from AGB stars, it is possible to perfectly reproduce the observed trends, and no extra light element primary process (LEPP) is needed \citep[but see][]{pra18}. The inclusion of $\nu$-driven winds in GCE simulation results in a strong overproduction of the abundances of the elements from Sr to Sn with respect to the observations.
\item For the heavier neutron-capture elements, contributions from both NS-NS/NS-BH mergers and MRSNe are necessary, and the latter is included in the blue areas.
Note again that the green areas includes the mass-loss component, i.e., not newly produced but recycled.
Note that Tc and Pm are radioactive.
\end{itemize}
\begin{figure}[t]
\begin{center}
  \includegraphics[width=0.75\textwidth]{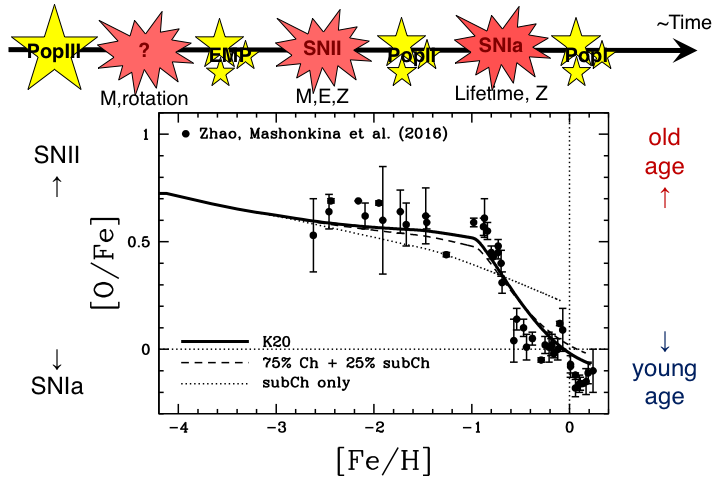}
\vspace*{-2mm}
\caption{The [O/Fe]--[Fe/H] relations in the solar neighborhood for the models with Ch-mass SNe Ia only (solid line), 75\% Ch plus 25\% sub-Ch-mass SNe Ia (dashed line), and sub-Ch-mass SNe Ia only (dotted line). The observational data (filled circles) are high-resolution non-local thermodynamic equilibrium (NLTE) abundances from \citet{zhao16}.
Figure is taken from \citet{kob20ia} with modification.
}
\label{fig:ofe}
\end{center}
\end{figure}

Since the Sun is slightly more metal-rich than the other stars in the solar neighborhood (see Fig.\,2 of K20), the fiducial model in K20 goes through [O/Fe]$=$[Fe/H]$=0$ slightly later compared with the Sun's age.
Thus, a slightly faster star formation timescale ($\tau_{\rm s}=4$ Gyr instead of 4.7 Gyr) is adopted in this figure.
The evolutionary tracks of [X/Fe] are almost identical.
In this model, the O and Fe abundances go though the cross of the dotted lines, meaning [O/Fe] $=$ [Fe/H] $=0$ at 4.6 Gyr ago.
This is also the case for some important elements including N, Ca, Cr, Mn, Ni, Zn, Eu, and Th. The remaining problems can be summarised as follows:
\begin{itemize}
\item
The contribution from rotating massive stars is not included in the K20 model, which can probably explain in the underproduction of C and F \citep{kob22iau}. A binary effect during stellar evolution may also increase C.
These elements could be enhanced by AGB stars as well, but the observed high F abundance in a distant galaxy strongly supported rapid production of F from Wolf-Rayet stars \citep[][\S \ref{sec:firstgal}]{franco21}.
\item 
Mg is slightly under-produced in the model, although at low metallicity the model [Mg/Fe] is slightly higher than observed (see Fig.\,\ref{fig:xfe-sagb}).
This may be due to a NLTE effect \citep{lind22}, or due to a binary effect. Massive stars in binaries tend to have a smaller CO core with a higher C/O ratios \citep{brown01}, which could result in a higher Mg/O ratio. Observed Mg/O ratios suggest that this binary effect should not be important at low metallicities ($\ltsim 0.1Z_\odot$).
\item
The underproduction of the elements around Ti is a long-standing problem since \citet{kob06}. It was shown that these elemental abundances can be enhanced by multi-dimensional effects of core-collapse supernovae (\citealt{sne16}; see also K15 model in K20). This is due to the lack of 3D simulations that can capture the production of these (less abundant) elements. 2D nucleosynthesis calculation showed an enhancement of these elements \citep{mae03,tom09}.
\item
The s-process elements are slightly overproduced even with the updated s-process yields.
Notably, Ag is over-produced by a factor of $6$, while Au is under-produced by a factor of $5$. U is also over-produced. These problems may require revising nuclear physics modelling namely fission \citep[see][for more discussion]{kob22iau}.
\end{itemize}

\begin{figure}[t]
\begin{center}
  \includegraphics[width=0.5\textwidth]{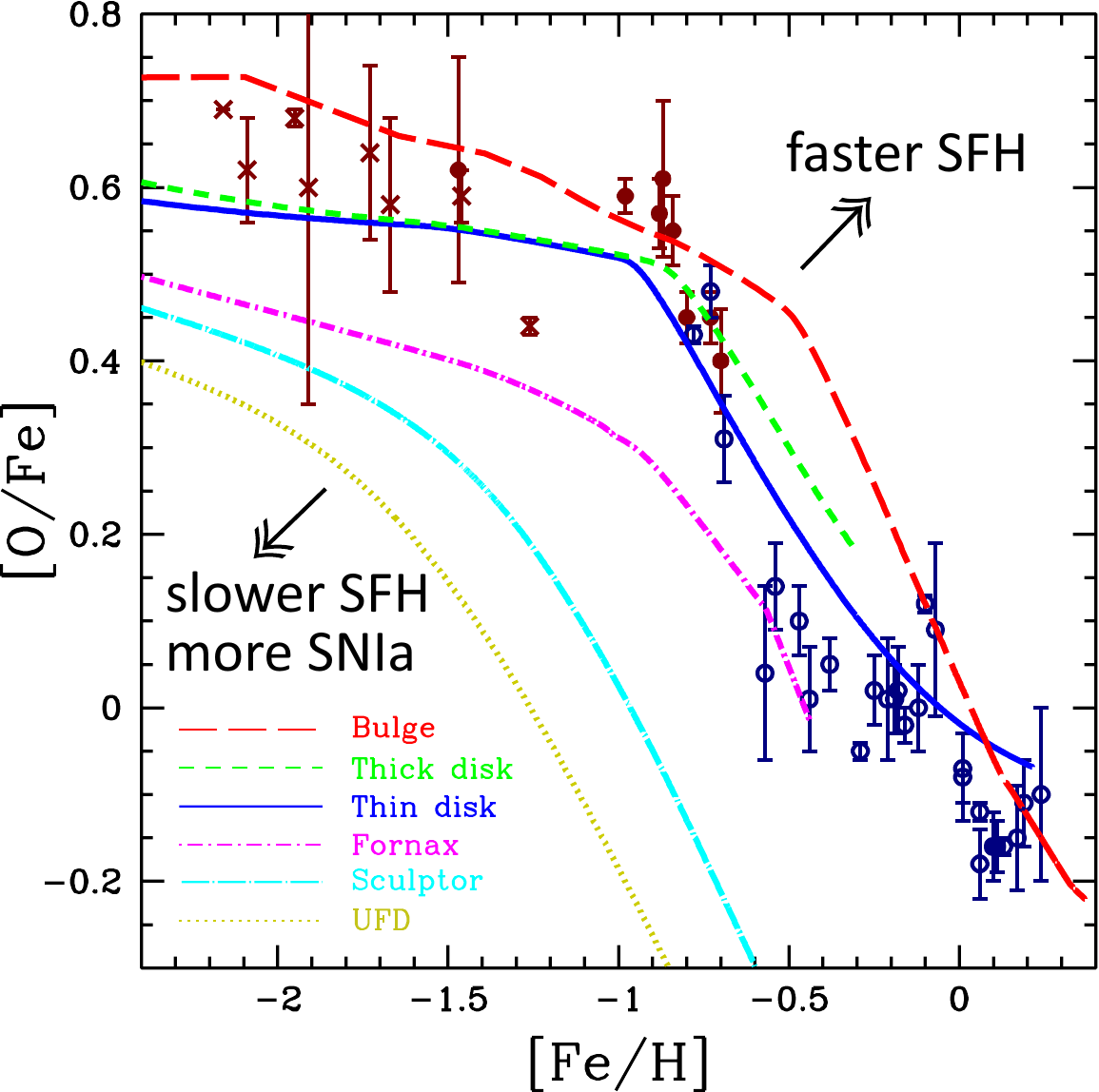}
\caption{The [O/Fe]--[Fe/H] relations in the Galactic bulge (red long-dashed line), thick disk (green short-dashed line), thin disk (blue solid line), Fornax (magenta dot-short-dashed line), Sculptor (cyan dot-long-dashed line), and ultra-faint dwarf (UFD) galaxies (olive dotted line). The star formation history of each system is constrained from its MDF in Figs.\,\ref{fig:mdf} and \ref{fig:mdf2}. See Fig.\,\ref{fig:ofe} for the observational data source for thin (navy open circles) and thick (maroon filled circles) disk stars. 
}
\label{fig:ofe-env}
\end{center}
\end{figure}

\subsection{The [X/Fe]--[Fe/H] relations}

Although the slopes in Figure \ref{fig:origin} show the different enrichment timescales of each element, it is easier to see the differences in the [X/Fe]--[Fe/H] diagrams\footnote{There were attempts to plot against a different element (e.g., O) to avoid the uncertainty of Fe yields \citep[e.g.,][]{cay04}, but it became rather hard to understand the plots.}.
Figure \ref{fig:ofe} shows the [$\alpha$/Fe]--[Fe/H] relation, which is probably the most important diagram in GCE; O is one of the $\alpha$ elements. 
At the beginning of the universe, the first stars (Population III stars) form and die, whose properties such as mass and rotation are uncertain and have been studied using the abundance patterns of the second generation, extremely metal-poor (EMP) stars.
Secondly, core-collapse supernovae occur, and their yields are imprinted in the abundance patterns of Population II stars in the Galactic halo. The [$\alpha$/Fe] ratio is high and stays roughly constant\footnote{There was a debate if the [O/Fe] ratio increases toward the lowest metallicity or not. UV OH line showed such an increase \citep[e.g,][]{isr98} reaching [O/Fe] $\sim 1$ at [Fe/H] $\sim -3$, which was later found to be due to 3D effects of stellar atmosphere.} with a small scatter. This plateau value does not depend on the star formation history but does on the IMF.
Finally, SNe Ia occur, which produce more Fe than $\alpha$ elements, and thus the [$\alpha$/Fe] ratio decreases toward higher metallicities; this decreasing trend is seen for the Population I stars in the Galactic disk.

The contribution from SNe Ia depends on the progenitor binary systems, namely, the mass of the progenitor WDs. Sub-Ch mass explosions produce less Mn and Ni, and more Si, S, and Ar than near-Ch mass explosions \citep{sei13,kob20ia}.
Figure \ref{fig:ofe} shows the [O/Fe]--[Fe/H] relations with varying the fraction of sub-Ch-mass SNe Ia. Including up to 25\% sub-Ch mass contribution to the GCE (dashed line) gives a similar relation as the K20 model (solid line), while the model with 100\% sub-Ch-mass SNe Ia (dotted line) gives too low an [O/Fe] ratio compared with the observational data. For Ch-mass SNe Ia, the progenitor model is based on the single-degenerate scenario with the metallicity effect due to optically thick winds from WDs \citep{kob98}. For sub-Ch-mass SNe Ia, the observed delay-time distribution is used since the progenitors are the combination of mergers of two WDs in double degenerate systems and low accretion in single degenerate systems; \citet{kob15}'s formula are for those in single degenerate systems only.

Because of this [$\alpha$/Fe]--[Fe/H] relation, high-$\alpha$ and low-$\alpha$ are often used as a proxy of old and young ages of stars in galaxies, respectively. Note that, however, that this relation is not linear but is a plateau and decreasing trend (called a `knee'). The location of the `knee' depends on the star formation timescale, with a higher metallicity for faster star formation history \citep{mat90}, e.g., [Fe/H] $\sim -0.5, -0.8, -1, -2$ for the Galactic bulge, thick disk, thin disk, and satellite galaxies.
Figure \ref{fig:ofe-env} shows the GCE model results for various environments. The bulge (with outflow), thick disk, and solar neighborhood (thin disk) models are taken from K20 with 100\% Ch-mass SNe Ia, with the star formation histories in Fig.\,\ref{fig:mdf}. The models for dwarf spheroidal galaxies are taken from \citet{kob20ia} adding 100\% sub-Ch mass SNe Ia on top of the 100\% of Ch mass SNe Ia, with the star formation histories in Fig.\,\ref{fig:mdf2}.
Recently, a similar relation is also shown for M31 using planetary nebulae. Since Fe is not available, Ar is used as a significant fraction of Ar is produced by SNe Ia. The thin and thick disk dichotomy seems to exist, but not exactly the same as in the Milky Way, probably due to M31 experiencing an accretion of a relatively massive, gas-rich satellite galaxy \citep{arnaboldi22}.

There are GCE models that try to explain both thin and thick disk observations with two infalls \citep{chi97,spitoni19}. However, more realistic, chemodynamical simulations show that a significant number of thick disk stars are formed in satellite galaxies before they accrete onto the disk, which can be better approximated with two independent GCE models in this figure \citep[also in][]{gri17}.

\begin{figure*}[t]
\center
\includegraphics[width=0.95\textwidth]{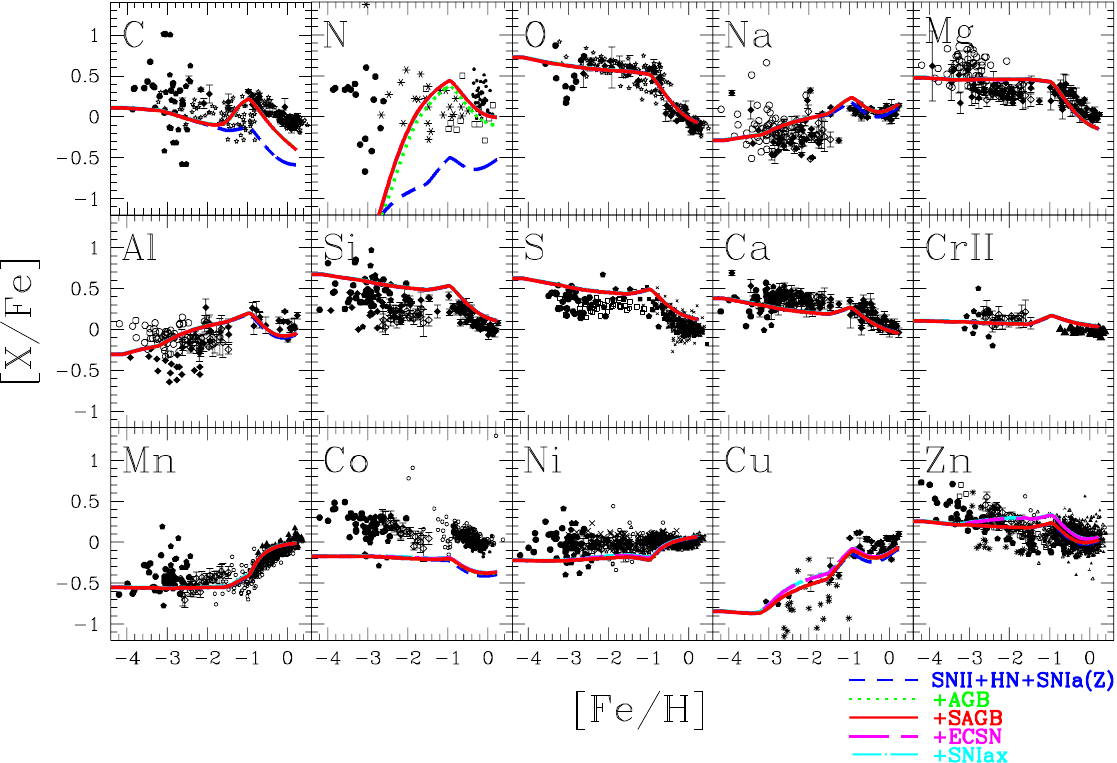}
\caption{\label{fig:xfe-sagb}
Evolution of the elemental abundances [X/Fe] from C to Zn against [Fe/H]
for the models in the solar neighborhood,
with only supernovae (without AGB and super-AGB stars, blue short-dashed line),
with AGB without super-AGB stars (green dotted lines),
with AGB and super-AGB stars (red solid line, fiducial model),
with ECSNe (magenta long-dashed lines),
and with SNe Iax (cyan dot-dashed lines).
Figure is taken from \citet{kob20sr} and see the reference for the observational data sources.
}
\end{figure*}

Figure \ref{fig:xfe-sagb} shows the evolution of elemental abundance ratios [X/Fe] against [Fe/H] from C to Zn in the solar neighborhood. 
All $\alpha$ elements (O, Mg, Si, S, and Ca) show the [$\alpha$/Fe]--[Fe/H] relations, i.e., the plateau and decreasing trend from [Fe/H] $\sim -1$ to higher metallicities.
The odd-$Z$ elements (Na, Al, and Cu) show an increasing trends toward higher metallicities due to the metallicity dependence of the core-collapse supernova yields.
See K20 for more detailed discussion on the evolutionary trends of each element, and \citet{kob22uv} for the discussion with more recent observations for Cu and Zn. In the following we focus the role of each enrichment source. 
\begin{itemize}
\item
The contribution to GCE from AGB stars (green dotted lines in Fig.\,\ref{fig:xfe-sagb}) can be seen mainly for C and N, and only slightly for Na, compared with the model that includes supernovae only (blue dashed lines).
Hence it seems not possible to explain the O--Na anti-correlation observed in globular cluster stars \citep[e.g.,][]{kraft97} with AGB stars \citep[but see][]{ventura09}.
Although AGB stars produce significant amounts of Mg isotopes, their inclusion does not affect the [Mg/Fe]--[Fe/H] relation.
\item
The contribution from super-AGB stars (red solid lines) is very small; with super-AGB stars, C abundances slightly decrease, while N abundances slightly increase.
It would be very difficult to put a constraint on super-AGB stars from the average evolutionary trends of elemental abundance ratios, but it might be possible to see some signatures of super-AGB stars in the scatters of elemental abundance ratios.
\item
With ECSNe (magenta long-dashed lines), Ni, Cu and Zn are slightly increased. These yields are in reasonable agreement with the high Ni/Fe ratio in the Crab Nebula \citep{nom87,wan09}.
\item
No difference is seen with/without SNe Iax (cyan dot-dashed lines) in the solar neighborhood because the progenitors are assumed to be hybrid WDs, which has a narrow mass range in the adopted super-AGB calculations \citet[][$\Delta M\sim0.1M_\odot$]{doh15}.
This mass range depends on convective overshooting, mass-loss, and reaction rates.
Even with the wider mass range in \citet[][$\Delta M\sim1M_\odot$]{kob15}, however, the SN Iax contribution is negligible in the solar neighborhood, but it can be important at lower metallicities such as in dwarf spheroidal galaxies with stochastic chemical enrichment \citep{ces17}.
\end{itemize}

\begin{figure*}[t]
\center
  \includegraphics[width=0.95\textwidth]{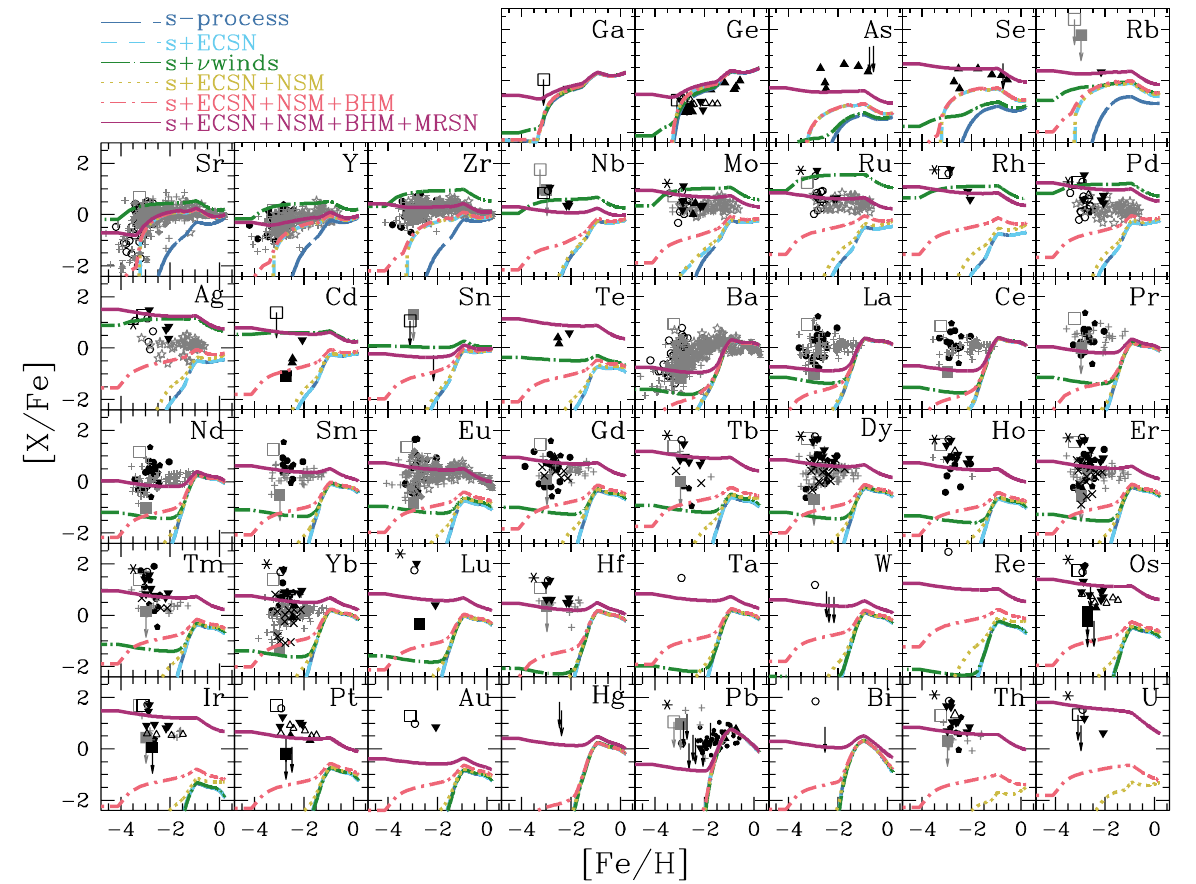}
\caption{The [X/Fe]--[Fe/H] relations for neutron capture elements, comparing to the models in the solar neighborhood, with s-process from AGB stars only (blue long-dashed lines), with s-process and ECSNe (light-blue short-dashed lines), with s-process, ECSNe, and $\nu$-driven winds (green dotted-long-dashed lines), with s-process, ECSNe, and NS-NS mergers (olive dotted lines), with s-process, ECSNe, and NS-NS/NS-BH mergers (orange dotted-short-dashed lines), with s-process, ECSNe, NS-NS/NS-BH mergers, and MRSNe (red solid lines). 
Figure is updated from \citet{kob20sr} with more observational data.
}
\label{fig:xfe}
\end{figure*}
\noindent
Figure \ref{fig:xfe} shows the evolutions of neutron-capture elements as [X/Fe]--[Fe/H] relations. 
\begin{itemize}
\item
As known, AGB stars can produce the first (Sr, Y, Zr), second (Ba), and third (Pb) peak s-process elements, but no heavier elements (navy long-dashed lines).
The second-peak elements are under-produced around [Fe/H] $\sim-2$, which is eased with chemodynamical simulations, and can also be reproduced better with rotating massive stars. However, the model with rotating massive stars results in over-production of the first peak elements \citep{kob22iau}.
\item
It is surprising that ECSNe from a narrow mass range ($\Delta M \sim 0.15-0.2M_\odot$) can produce enough of the first peak elements; with the combination of AGB stars, it is possible to reproduce the observational data very well (cyan short-dashed lines). This means that no other light element primary process (LEPP), such as rotating massive stars, is required \citep[but see][]{ces14}. The elements from Mo to Ag seem to be overproduced, which could be tested with the UV spectrograph proposed for the VLT, CUBES.
\item
Additional production from $\nu$-driven winds leads to further over-production of these elements in the model (green dot-long-dashed lines), but this should be studied with more self-consistent calculations of supernova explosions.
\item
Neutron star mergers can produce lanthanides and actinides, but not enough (olive dotted lines); the rate is too low and the timescale is too long, according to binary population synthesis. This is not improved enough even if NS-BH mergers are included (orange dot-short-dashed lines).
It is very unlikely this problem will be solved even if the mass-dependence of the nucleosynthesis yields are included, unless the BH spins are unexpectedly high \citep{kob22}.
\item
In the GCE model with MRSNe (magenta solid lines), it is possible to reproduce a plateau at low metallicities for Eu, Pt, and Th, relative to Fe. However, even with including both MRSNe and neutron star mergers, the predicted Au abundance is more than ten times lower than observed. This underproduction is seen not only for the solar abundance but also for low metallicity stars although the observational data are very limited. UV spectroscopy with HST, or NASA's future LUVOIR, is needed for investigating this problem further.
\end{itemize}

In conclusion, an r-process associated with core-collapse supernovae, such as MRSNe, is required. The same conclusion is obtained with other GCE models and more sophisticated chemodynamical simulations \citep[e.g.,][]{hay19,van20}, as well as from the observational constraints of radioactive nuclei in the solar system \citep{wal21}. See more discussion in the later section.

It seems not to be easy to solve the missing gold problem with astrophysics; only Au yields should be increased since Pt is already in good agreement with the current model, and Ag is rather overproduced in the current model. However, there are uncertainties in nuclear physics, namely in some nuclear reaction rates and in the modelling of fission,
which might be able to increase Au yields only, without increasing Pt or Ag.
It may be hard to predict the only one stable isotope of $^{197}$Au, while Pt has several stable isotopes.
The predicted Th and U abundances are after the long-term decays, to be compared with observations of metal-poor stars, and the current model does not reproduce the Th/U ratio either.

\begin{figure}[t]
\begin{center}
\includegraphics[width=0.7\textwidth]{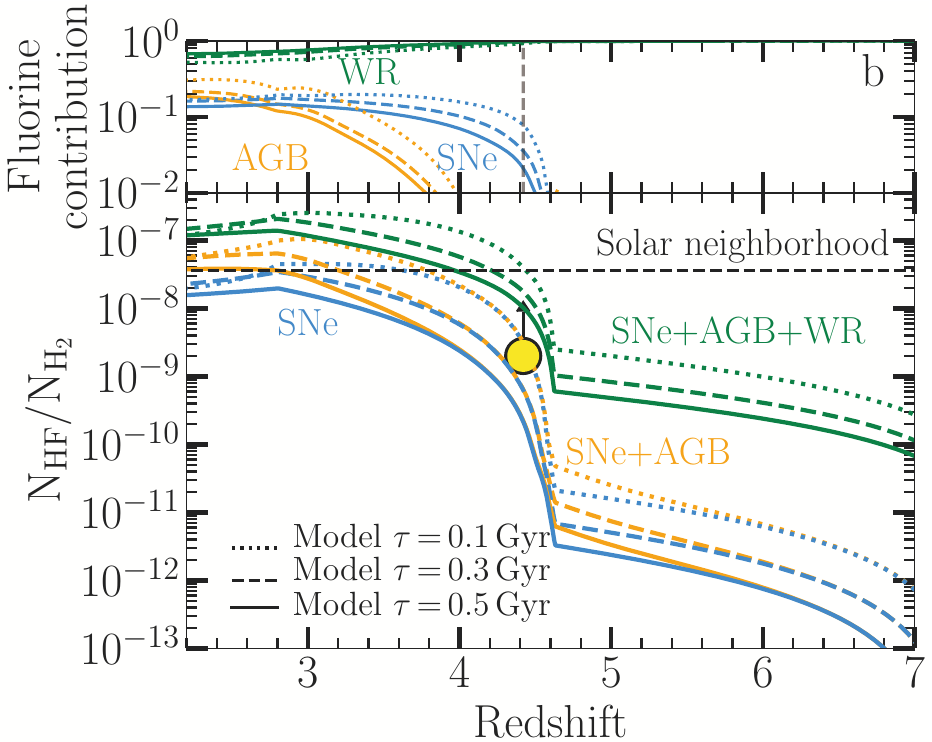}
\caption{The fluorine abundance in the galaxy NGP-190387 at redshift $z=4.4$ (yellow point), comparing with GCE models with only supernovae (blue lines), plus AGB stars (orange lines), plus WR stars (green lines).
Figure is taken from \citet{franco21}.
}
\label{fig:franco}
\end{center}
\end{figure}

\begin{figure}[t]
\begin{center}
\includegraphics[width=0.65\textwidth]{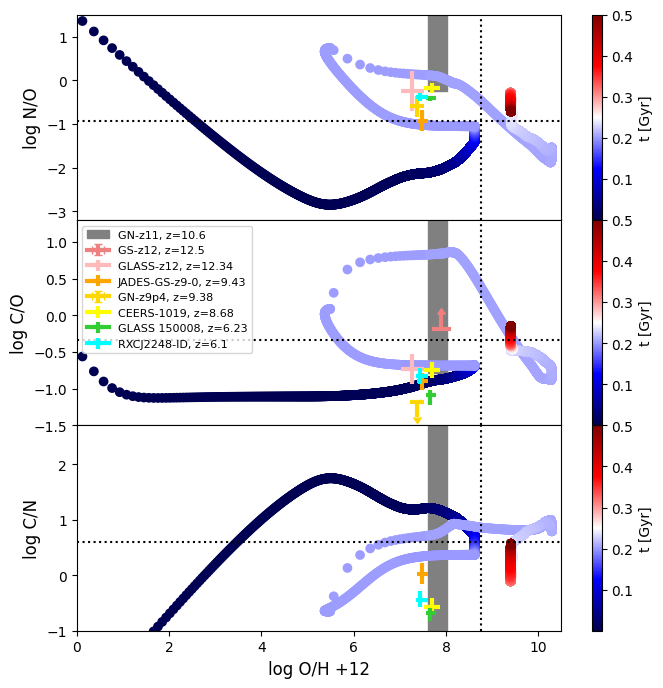}
\vspace*{-1mm}
\caption{\label{fig:gnz11}
Evolution of CNO (number) abundances in GCE models for the galaxy GN-z11 at $z=10.6$, assuming dual starburst with standard IMF, color-coded with the time (in Gyr) elapsed since the first star formation episode. The dotted lines indicate the solar ratios.
The gray bars show the observational data \citep{cameron23}.
Figure is adapted from \citet{kob24}.
}
\end{center}
\end{figure}

\subsection{The first galaxies}
\label{sec:firstgal}

The JWST was built to detect the first galaxies made of the first stars. As predicted by theorists, the timescale of chemical enrichment from the first stars is so short ($\sim2$ Myrs) that the JWST has not found a metal-free galaxy yet. Very high-redshift galaxies however may still possess the information on the first stars.
The Atacama Large Millimeter/submillimeter Array (ALMA) has opened a new window to study elemental abundances and isotopic ratios of light elements at very high redshifts \citep[e.g.,][]{zhang18,franco21}, which provide us an independent constraint on stellar nucleosynthesis.
This is rapidly expanded by the JWST. 
It is not so surprising to see the high N/O ratio in super-early galaxies such as GN-z11 \citep[$z=10.6$][]{bunker23}.
In the early Universe, stellar rotation may be important at low metallicities because weaker stellar winds result in smaller angular momentum loss.

The impact of stellar rotation on stellar evolution and nucleosynthesis has been studied to explain Wolf-Rayet (WR) stars (e.g., \citealt{mey02,hir07,lim18}), and the observed N and $^{13}$C abundances at low metallicities have been used to constrain the effect (\citealt{chi06}; see also Fig.\,13 of \citealt{kob11agb}).
With rotation, both C and N yields are increased due to the interplay between the core He-burning and the H-burning shell, triggered by the rotation-induced instabilities \citep[\S \ref{sec:sn}]{lim18}.
As discussed in the previous section, the importance of jet-like supernova explosion for iron-peak elements (Co and Zn) and r-process elements (via MRSNe) also indicates the importance of stellar rotation, although how to transport the angular momentum in stellar envelopes to stellar cores is uncertain.

Fluorine is another element enhanced by stellar rotation;
high abundance of CNO enhances the production of F in He convective shell \citep[see also][for other processes]{kob11f}.
This element is difficult to measure in stellar spectra;  accessible lines are only in infrared, and thus the sample number was limited \citep[e.g.,][]{jorissen92,cunha03,lucatello11}. There was also confusion in the excitation energies and transition probabilities for the HF lines \citep{jonsson14}. The vibrational-rotational lines get too weak at low metallicities, and the available sample of EMP stars is only for carbon enhanced stars \citep{mura20}.
However, it turns out that it is quite easy to measure the fluorine abundance in gas with ALMA.
The galaxy NGP-190387 was discovered by the H-ATLAS survey with the Herschel satellite, with the redshift confirmed by the Northern Extended Millimetre Array (NOEMA). It is a gravitationally lensed galaxy with a magnification factor $\mu\simeq5$. The lowest rotational transitions from the HF molecule appeared as an absorption line in the ALMA data. This molecule is very stable and the dominant gas-phase form of fluorine in the ISM.
Figure \ref{fig:franco} shows the fluorine abundance of NGP-190387 (yellow point), comparing with GCE models with various star formation timescales. 

These models used \citet{lim18}'s yields (set R) assuming the rotational velocity of 300 km s$^{-1}$ for 13--120$M_\odot$ stars.
It is also assumed that star formation took place soon after the reionization, which was boosted probably due to galaxy merger 100 Myr before the observed redshift $z=4.4$. This assumption is based on the properties of other submillimetre galaxies at similar redshifts. The models show rapid chemical enrichment including fluorine, but in order to explain the observed fluorine abundance, the contribution from WR stars (green lines) is required.

The evolution of isotopic ratios is also largely affected by stellar rotation. Without rotation, $^{13}$C and $^{25,26}$Mg are produced from AGB stars ($^{17}$O might be overproduced in our AGB models), while other minor isotopes are more produced from metal-rich massive stars, and thus the ratios between major and minor isotopes (e.g., $^{12}$C/$^{13}$C, $^{16}$O/$^{17,18}$O) generally decrease as a function of time/metallicity (Figs.\,17--19 of \citet{kob11agb} and Fig.\,31 of K20; see also \citealt{rom19}).
Isotopic ratios of CNO, Si, S, Cl, and Ar have been estimated for a couple of spiral galaxies around $z\sim1$ \citep[][and the references therein]{wallstrom19}, as well as for various sources in the Milky Way and nearby galaxies.

With rotation, $^{13}$C and $^{17,18}$O are enhanced. The evolution of isotopic ratios and comparison to observations can be found in Fig.\,7 of \citet{kob22iau}, which gives consistent results with \citet{rom19}'s GCE models. $^{12}$C/$^{13}$C ratio became too high, while it is possible to explain the low $^{13}$C/$^{18}$O ratio observed in submillimetre galaxies at redshift $z\sim2$--3 \citep{zhang18} without changing the IMF.
$^{25,26}$Mg are also enhanced in \citet{lim18}; freshly-made $^{14}$N that diffused back to the center quickly converted into $^{22}$Ne and then into $^{25,26}$Mg.
Mg isotopic ratios are measurable with very high-resolution spectra of nearby stars \citep[e.g,][]{boesgaard68,yon03,carlos18}.
Comparing the latest observations to GCE models, \citet{mckenzie24} found that $^{26}$Mg is overproduced in \citet{lim18}'s yields.
It is important to constrain the rotational induced mixing using these nuclei that are formed in different layers of massive stars.

The galaxy GN-z11 was one of the most distant galaxy candidate found with HST by \citet{oesch16}. JWST/NIRSpec showed an amazing spectrum with strong N emission lines at $z=10.6$, only 430 Myrs after the Big Bang \citep{bunker23}.
N production from WR stars is not sufficient to explain the observation. More specifically, it is possible to reproduce the observed $\log$ N/O $> -0.25$ at a much lower metallicity, but not at the observed $\log$ O/H $=7.82$, even with an extremely short star formation timescale or changing the IMF.
\citet{kob24} solved this assuming an intermittent star formation.
Figure \ref{fig:gnz11} shows our GCE model where N is enhanced from the second star burst from WR stars, before the bulk of oxygen is produced from core-collapse supernovae.
Such intermittent star formation is common in cosmological simulations, and should also be in the early Universe.

Several more N-rich galaxies have been reported, but there seems to be a variety in the C-abundances, which can be explain by the IMF variation \citep{curti24,arellano24}.
Although N abundance is not measured (only upper limits), a carbon-rich galaxy with $[{\rm C/O}]>0.15$ at $z=12.5$, only 350 Myrs after the Big Bang, is also reported \citep{deugenio24}.

In our new GCE models with WR stars, the C and N yields from \citet{lim18} are combined with the K20 yields.
This is because \citet{lim18}'s yields do not reproduce the observations of iron-peak elements either, because of their simple description of supernova explosions.
C is also overproduced \citep{rom19}, which is a serious problem for constraining the IMF from observed C/N ratios in distant galaxies.
Our GCE models with WR stars are calibrated to reproduce the observed [(C,N,O)/Fe] abundances in the Milky Way.
See also Fig.\,8 of \citet{kob22iau}.

\subsection{The first chemical enrichment}
\label{sec:first}

\begin{figure}[t]
\begin{center}
\includegraphics[width=0.65\textwidth]{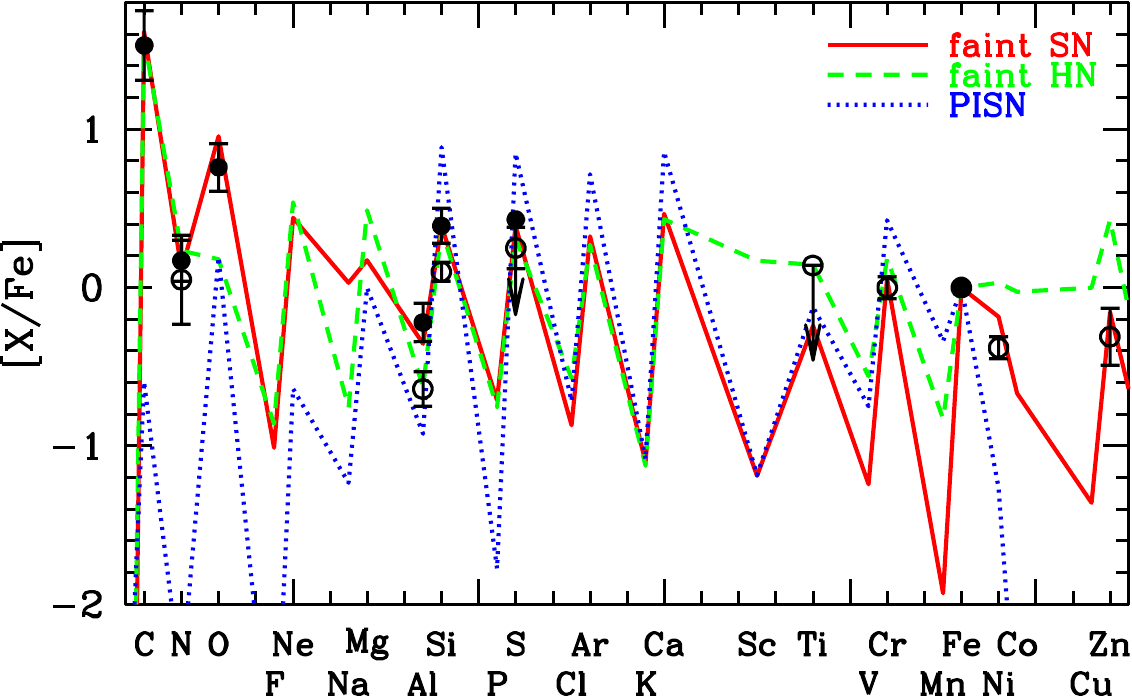}
\caption{\label{fig:cdla}
Elemental abundance patterns of the Population III supernovae.
The red solid and green short-dashed lines show the nucleosynthesis yields of faint core-collapse supernovae from $25M_\odot$ stars with mixing fallback. The blue dotted line is for pair-instability supernovae from $170M_\odot$ stars.
Stellar rotation is not included.
The dots are for the metal-poor C-rich DLA (filled circles) and peculiar DLA (open circles). 
Figure is taken from \citet{kob11dla} with modification.}
\end{center}
\end{figure}

Although the first stars can be massive (\S \ref{sec:vms}), it would not be possible observe even with the JWST. In the Milky Way, no metal-free star is found, and the most metal-poor star has [Fe/H] $\ltsim -7$ and is carbon enhanced \citep[SMSS0313-6708, ][]{kel14,nor17}.
Extremely metal-poor (EMP) stars ([Fe/H] $<-3$, \citealt{bee05}) have been an extremely useful relic in the Galactic archaeology. At the beginning of galaxy formation, stars form from a gas cloud that was enriched only by one or very small number of supernovae \citep{audouze95}, and hence the elemental abundances of EMP stars can offer observational evidence of particular types of supernovae in the past.
The expectation was that the first stars were so massive ($\sim 140$--$300M_\odot$) that they exploded as a pair-instability supernova \citep[PISN,][]{barkat67}, which causes high [(Si,S)/O] ratios \citep[e.g.,][]{nom13}.

However, after half a century of surveys, no star has been found with an elemental abundance pattern fitted by a PISN in the Milky Way \citep{sku24}.
Instead, it is found that quite a large fraction of massive stars become faint supernovae \citep{ume03} that give a high C/Fe ratio leaving a relatively large black hole ($\sim 5M_\odot$). 
If there were C-enhanced low-$\alpha$ stars, that would indicate black hole formation even from 10--20$M_\odot$ Pop III stars \citep{kob14}.
From faint supernovae, the ejected iron mass is very small but the explosion energy can be high; faint hypernovae can explain the Zn enhancement of CEMP stars \citep{ish18}.

The C-enhancement can also be caused by binary mass transfer, and many of the CEMP stars with s-process enhancement are binaries \citep[CEMP-s][]{han16}.
If there were faint supernovae and CEMP-no stars formed from the C-enhanced ejecta, then there must be gas with a similar C-enhancement in the past.
\citet{kob11dla} expanded `abundance profiling' to quasar absorption line systems, and showed that the C enhancement of a damped Ly-$\alpha$ system (DLA) at $z=2.34$ with [Fe/H] $\simeq -3$ is very similar to the CEMP stars at [Fe/H] $=-4$.
This observation is found to be incorrect later, but recently similar C enhancement is shown for a few sub-DLAs \citep{saccardi23}.
Figure \ref{fig:cdla} shows the comparison between the nucleosynthesis yields of a faint supernova (red solid line), a faint hypernova (green dashed line), and a PISN (blue dotted line).

\begin{figure}[t]
\begin{center}
  \includegraphics[width=0.65\textwidth]{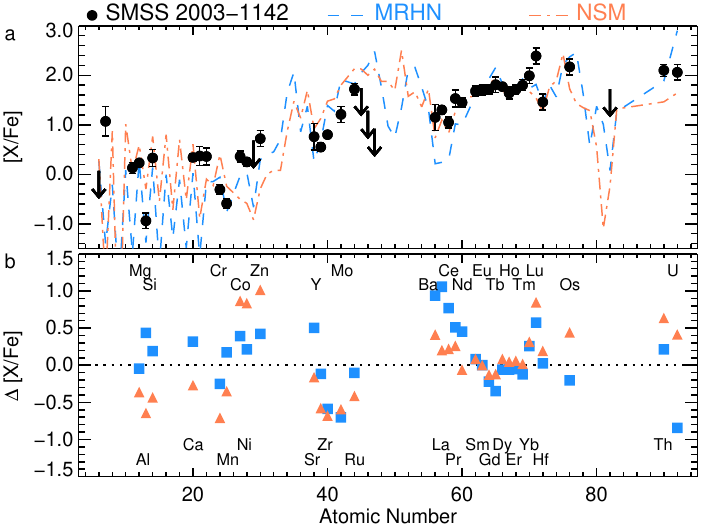}
\caption{The elemental abundance of an extremely metal-poor star SMSS J200322.54-114203.3, which has [Fe/H]$= -3.5$, comparing with mono-enrichment from a magneto-rotational hypernovae (blue dashed line and squares) and multi-enrichment including a neutron star merger (orange dot-dashed line and triangles).
The lower panel shows the differences, i.e., the observed values minus model values.
Figure is taken from \citet{yon21a}.}
\label{fig:yong}
\end{center}
\end{figure}

A small number of EMP stars show a relatively low $\alpha$ abundance, which does not necessarily mean that the ISM from which the star formed was already enriched by SNe Ia. 
The reasons that could cause low $\alpha$/Fe ratios were summarised in \citet{kob14}: (1) SNe Ia, (2) less-massive core-collapse supernovae ($\ltsim 20M_\odot$), which become more important with a low star formation rate, (3) hypernovae, although the majority of hypernovae are expected to give normal [$\alpha$/Fe] ratios ($\sim 0.4$), and (4) PISNe, which could be very important in the early Universe.
Therefore, the [$\alpha$/Fe] ratio is not a perfect clock.
It is necessary to also use other elemental abundances (namely, Mn and neutron-capture elements) or isotopic ratios, with higher resolution ($>40,000$) multi-object spectroscopy on 8m class telescopes e.g., WFMOS (cancelled in 2009), MSE (suspended in 2024), and HRMOS.

EMP stars with r-process enhancement (called rII stars), namely those with enhancement of the heaviest elements (actinide-boost stars), can offer confirmation of an r-process site.
Recently, from the SkyMapper survey, \citet{yon21a} found such a star with [Fe/H] $=-3.5$, the lowest metallicity known for rII stars. The abundance pattern is presented in Figure \ref{fig:yong}, which shows a clear detection of U and Th, the universal r-process pattern, normal $\alpha$ enhancement, but unexpectedly showing high Zn and N abundances ([Zn/Fe]$=0.72$, [N/Fe]$=1.07$).
This abundance pattern cannot be explained with a model with a neutron star merger (orange dot-dashed line), but instead strongly supports a magnetorotational hypernova from a $\gtsim25M_\odot$ Pop III star (blue dashed line).

There are a few magneto-hydrodynamical simulations and post-process nucleosynthesis that successfully showed enough neutron-rich ejecta to produce the 3rd peak r-process elements as in the Sun \citep{win12,mos18}. The predicted iron mass is rather small \citep{nis15,rei21}, and the astronomical object may be faint. \citet{yon21a}'s proposal is Fe-producing, magnetorotational hypernova, which is more luminous.
The r-process elements could also be formed in the accretion disk of a collapsar \citep{sie19}, but the inner ejecta that contains Fe and Zn must be ejected at the explosion as for a hypernova. As it is luminous it might be possible to detect absorption features of Au and Pt in the spectrum of a transient in future.

\begin{figure}[t]
\begin{center}
\includegraphics[width=11cm]{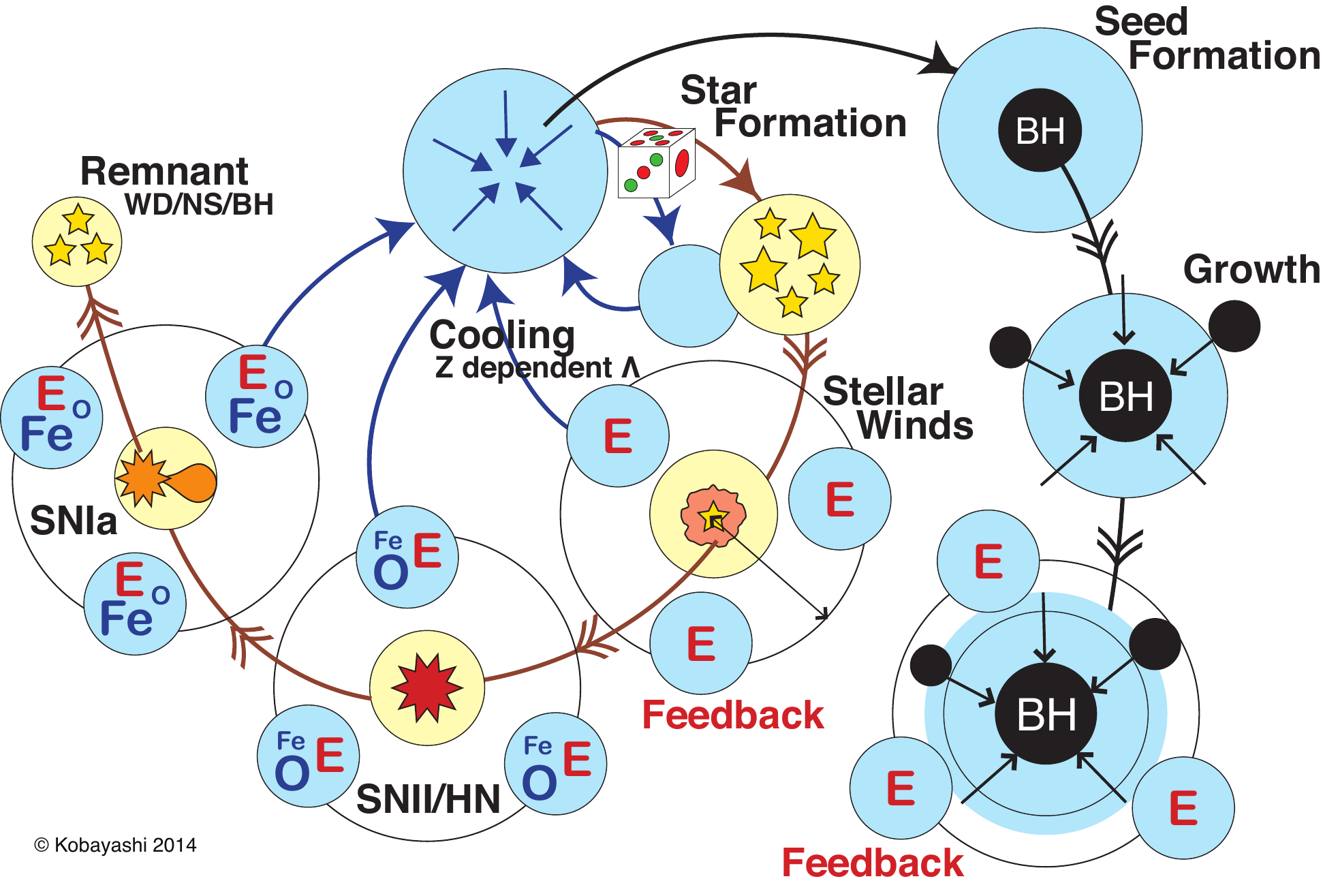}
\caption{\label{fig:hydro}
Schematic view of chemodynamical evolution of galaxies and super-massive black holes (BHs). $E$ (red) denotes energy, while O and Fe are elements.
}
\end{center}
\end{figure}

\section{Chemodynamical evolution of galaxies}
\label{sec:hydro}

Thanks to the development of super computers and numerical techniques, it is now possible to simulate the formation and evolution of galaxies from cosmological initial conditions, i.e., initial perturbation based on $\Lambda$ cold dark matter (CDM) cosmology. Star formation, gas inflow, and outflow in Eq.(\ref{eq:gce}) are not assumed but are, in principle, given by dynamics. The baryon cycle is schematically shown in Figure \ref{fig:hydro}. Due to the limited resolution, star forming regions, supernova ejecta, and active galactic nuclei (AGN) cannot be resolved in galaxy simulations, and thus it is necessary to model star formation and the subsequent effects -- feedback -- introducing a few parameters. Fortunately, there are many observational constraints, from which it is usually possible to choose a best set of parameters. To ensure this, it is necessary to run the simulation until the present-day, $z=0$, and reproduce a number of observed relations at various redshifts.

Although hydrodynamics can be calculated with publicly available codes such as Gadget, RAMSES, and AREPO, modelling of baryon physics is the key and is different depending on the simulation teams/runs.
There are various simulations of a cosmological box with periodic boundary conditions, and containing galaxies with a wide mass range (e.g., $10^{9-12} M_\odot$ in stellar mass) at $z=0$
\citep{sch15,dub16,dolag17,pillepich18a,dave19}.
In order to study massive galaxies it is necessary to increase the size of the simulation box (e.g., $\gtsim 100$ Mpc), while in order to study internal structures it is necessary to improve the spatial resolution (e.g., $\ltsim 0.5$ kpc).
The box size and resolution are chosen depending on available computational resources.
On the other hand, zoom-in techniques allow us to increase the resolution focusing a particular galaxy,
but this also requires tuning the parameters with the same resolution, comparing to a number of observations in the Milky Way.
Needless to say, elemental abundances are the most informative. There are a few zoom-in simulations for Milky Way-type galaxies \citep[e.g.,][]{brook12,few14,grand17,hop18fire,buck20,font20}, but in most of the cases the input nuclear astrophysics are too simple to make use of elemental abundances.

In these simulations all relevant baryon physics are included: radiative cooling, star formation, supernova feedback, but with different formula (see \citealt{kob23book} for more details).
For cosmological simulations, the formation, growth, and feedback of super-massive black holes (SMBHs) are also included, again with different formulae.
Without this AGN feedback it is not possible to reproduce 1) luminosity/mass functions of galaxies, and 2) early-type galaxies as observed.

For chemical enrichment,
\citet{kob04} introduced a scheme to follow the evolution of the star particle as a simple stellar population (SSP), which is defined as a single generation of coeval and chemically homogeneous stars of various masses, i.e. it consists of a number of stars with various masses but the same age and chemical composition.
The production of each element $i$ from the star particle (with an initial mass of $m_*^0$) is calculated using a very similar equation as Eq.(\ref{eq:gce}):
\begin{equation}\label{eq:e-z}
E_{Z_i}(t_{\rm birth}, Z_i, {\rm IMF})= m_*^0 \left[ E_{\rm SW}+E_{\rm SNcc}+E_{\rm SNIa}+E_{\rm NSM} \right] .
\end{equation}
With this SSP method, the identical equations for GCE, Eqs.\,(\ref{eq:e_sw}--\ref{eq:e_nsm}) can be used for the ejection rates $E_{\rm SW}$, $E_{\rm SNcc}$, $E_{\rm Ia}$, and $E_{\rm NSM}$.

Similarly, the energy production from the star particle is:
\begin{equation}\label{eq:e-e}
E_e(t) = m_*^0 \left[ e_{e,{\rm SW}}{\cal{R}}_{\rm SW}(t)+e_{e,{\rm SNcc}}{\cal{R}}_{\rm SNcc}(t)+e_{e,{\rm SNIa}}{\cal{R}}_{\rm SNIa}(t) \right] ,
\end{equation}
where the number of stars that start to cause SWs (${\cal{R}}_{\rm SW}$) and core-collapse supernovae (${\cal{R}}_{\rm SNcc}$) can be approximately calculated as
\begin{equation}
{\cal R}_{\rm SW}=\int_{\max[8M_\odot,\,m_t]}^{m_u}\,
\frac{1}{m}\,\psi(t-\tau_m)\,\phi(m)~dm ,
\end{equation}
\begin{equation}
{\cal R}_{\rm SNcc}=\int_{\max[8M_\odot,\,m_t]}^{\min[50M_\odot,\,m_u]}\,
\frac{1}{m}\,\psi(t-\tau_m)\,\phi(m)~dm .
\end{equation}
The total energy due to SWs per star (in erg) is $e_{e,{\rm SW}}=0.2 \times 10^{51} ({Z}/{Z_\odot})^{0.8}$ for $>8M_\odot$ stars.
The explosion energy of core-collapse supernovae $e_{e,{\rm SNcc}}$ is calculated from $e_{e,{\rm SNII}}=1 \times 10^{51}$ erg for Type II supernovae and $e_{e,{\rm HN}}=10,10,20,30 \times 10^{51}$ erg for $20,25,30,40 M_\odot$ hypernovae.
For binaries, the energy production from SNe Ia is significant and is $e_{e,{\rm SNIa}}=1.3 \times 10^{51}$ erg, while the energy production from NSMs is negligible.
The SNIa rate ${\cal{R}}_{\rm SNIs}$ is given by Eq.(\ref{eq:r_ia}).

\begin{figure}[t]
\center
\url{https://star.herts.ac.uk/~chiaki/works/diskold160000z.mpg}
\caption{
Time evolution of the formation of an isolated disk galaxy in a halo of mass of $10^{10} h^{-1}M_\odot$. The white points show star particles, while the gas particles are colour-coded according to their temperature.
The left panel shows face-on projections, the right panel gives edge-on views.
Each panel is $40$ kpc on a side. 
[Credit: \citet{kob07}]
}
\label{fig:diskmap}
\end{figure}

Movie \ref{fig:diskmap} shows galactic winds driven by supernova feedback in an isolated disk galaxy.
The initial condition is a rotating gas cloud with total mass of $10^{10}h^{-1}M_\odot$, and non-isotropic infall of gas form a dense disc, where stars form. As a result of the energy input, the low-density hot-gas region expands in a bipolar flow ($t\sim0.14$ Gyr) ejecting metals outside the galaxy. In the disk plane, the hot gas region expands and forms a dense shell where stars keep on forming. The energy from these stars can quickly propagate to the surrounding low-density region. After the galactic wind forms ($t\sim0.28$), the gas density becomes so low at the centre that star formation is terminated for a while. Because of radiative cooling, a part of the ejected, metal-enhanced gas, however, returns, again settling in the disk where it fuels new star formation, but this secondary star formation is not as strong as the initial starburst. Although some small bubbles are forming in the galaxy in this stage, not much gas and metals are ejected by them from the disk ($t\sim0.98$).
The galactic winds become much weaker in more massive galaxies (\S \ref{sec:cosmo}). 

\subsection{Galactic archaeology}
\label{sec:mw}

\begin{figure}[t]
\centering
\url{https://star.herts.ac.uk/~chiaki/works/Aq-C-5-kro2.mpg}
\caption{V-band luminosity map for the edge-on view of 200kpc (left panel) and 20kpc (right panel) on a side,
of our chemodynamical zoom-in simulation of a Milky Way-type galaxy.}
\label{fig:mwmap}
\end{figure}

In the Local Group, i.e., in the Milky Way and dwarf spheroidal (dSph) galaxies, the spatial distribution of detailed elemental abundances can be obtained from high-resolution spectra of individual low-mass stars\footnote{Ages of stars can also be well estimated with asteroseismology recently.}. This has been done for more than half a century, and the average trends of the observations for many elements have been used to constrain the stellar physics \citep{tim95,kob06,rom10}. The recent Galactic archaeology surveys with medium resolution MOS dramatically increased the sample and made possible to discuss the distributions of stars along the average trends, which requires more realistic, chemodynamical simulations of galaxies. Meantime, observations with different lines for each element \citep[e.g.,][]{isr98,sne16} revealed NLTE and 3D effects in the stellar atmosphere. Detailed star-by-star analysis of wide range of high-resolution spectra (including UV) is still required to obtain absolute values of elemental abundances \cite[][for the need of UV spectra]{kob22iau}.

Although there were simulations of isolated galaxies, \citet{kob11mw} was the first chemodynamical simulations that showed the evolution elements from O to Zn in a Milky Way-type galaxy from cosmological initial condition.
Our new simulation, which is run by a Gadget-3 based code (Kobayashi, in prep.), includes all stable elements, and a fully cosmological initial condition is applied \citep{sca12}.
The spatial and mass resolutions are 0.5 kpc and $3\times10^5M_\odot$, respectively.

Movie \ref{fig:mwmap} shows the time evolution of the edge-on views for V-band luminosities of star particles.
The galactic bulge formed by an initial star burst, by an assembly of gas-rich sub-galaxies beyond $z=2$.
The disk formed with a longer timescale and has grown {\it inside-out}; the disk was small in the past and becomes larger at later times.
In the $\Lambda$CDM cosmology, satellite galaxies keep accreting, which interact and merge with the main body, but there is no major merger after $z=2$. Otherwise, it is not possible to keep the disk structure as in the Milky Way. 
These galaxy mergers also make the disk thicker.
Approximately one third of thick disk stars already formed in merging galaxies, which are disrupted by tidal force and accreted onto the disk plane.
In the thin disk, star formation is self-regulated, and chemical enrichment takes place following cosmological gas accretion, radial flows, and stellar migrations; these physical processes are analysed in detail in \citet{vin20}.

\subsubsection{The $\alpha$/Fe bimodality}
\begin{figure}[t]
\begin{center}
  \includegraphics[width=0.48\textwidth]{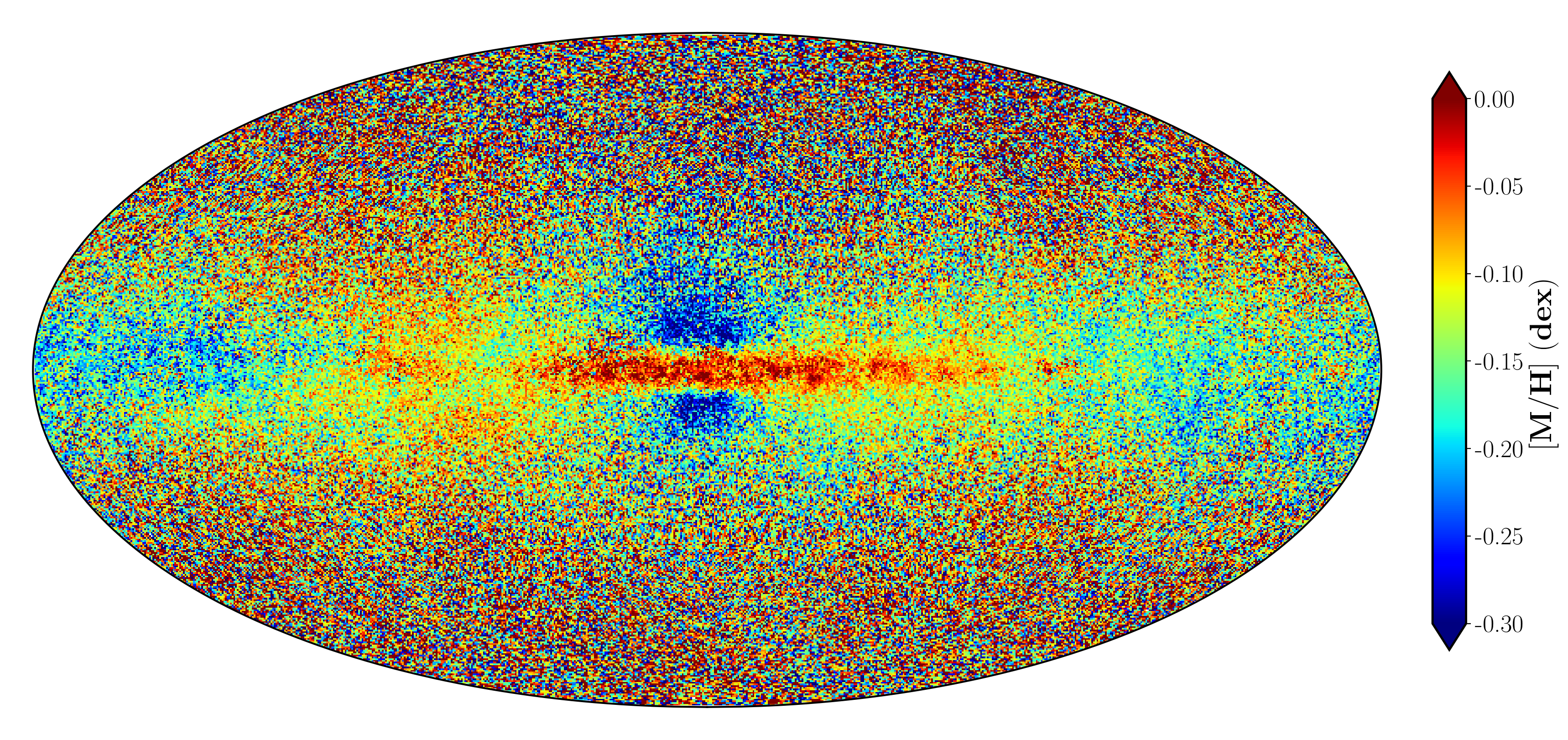}
  \includegraphics[width=0.48\textwidth]{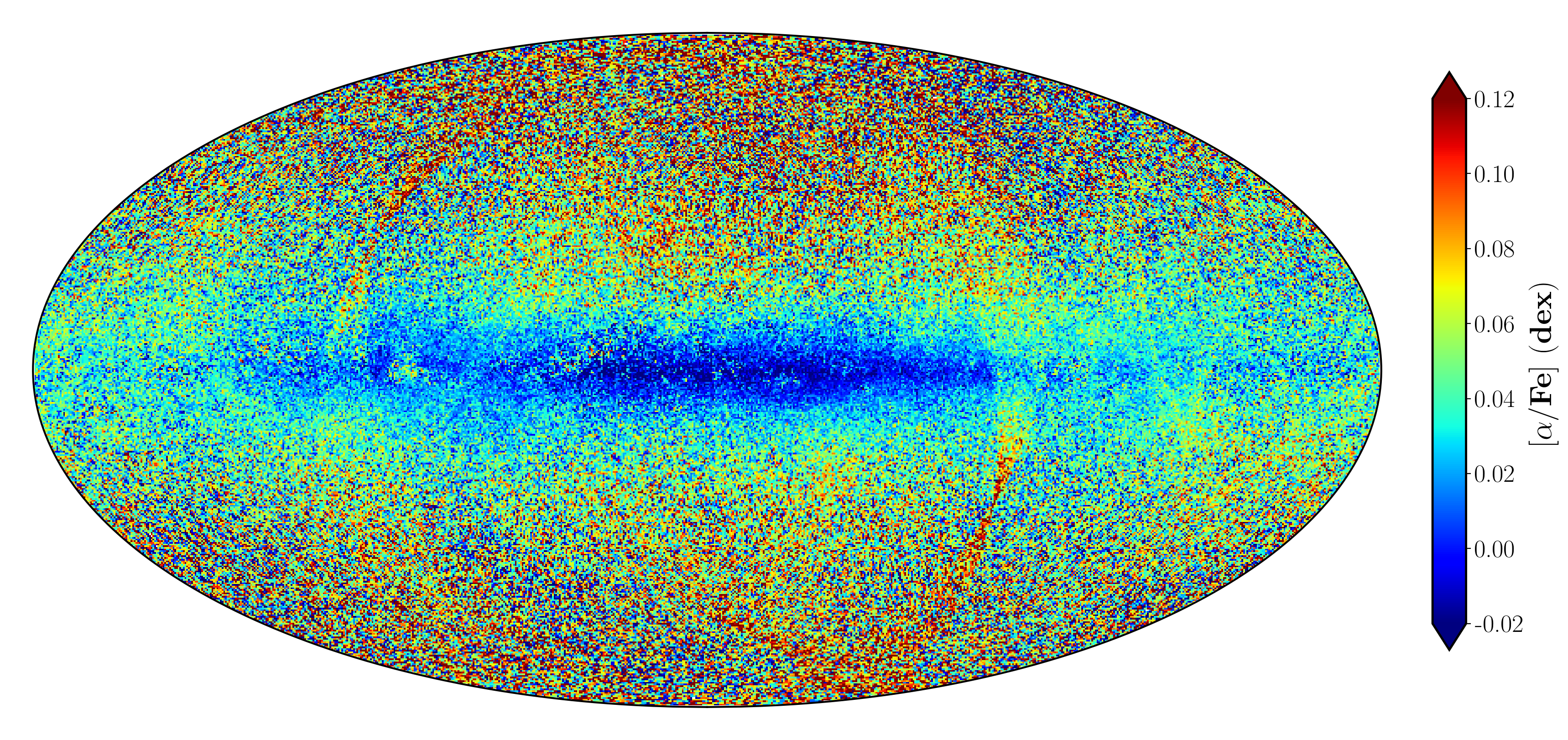}
\caption{(Left panel) Metallicity map of Gaia DR3 released on 13th June 2022, in the range of [M/H]$=-0.3$ (blue) to 0.0 (red).
(Right panel) The same as the left panel but for [$\alpha$/Fe]$=-0.02$ (blue) to 0.12 (red). Figures are taken from \citet{gaiadr3}.
}
\label{fig:gaia}
\end{center}
\end{figure}

At the present-day ($z=0$), our simulated galaxy shows very similar maps of metallicity and [$\alpha$/Fe] ratios as observed. Figure \ref{fig:gaia} shows the observations in Gaia DR3, which contains a half billion of stars. The data show too blue outer bulge in the metallicity map, and in the [$\alpha$/Fe] map the patterns close to the Ecliptic Poles are artifacts.
There are both vertical and radial metallicity gradients; metallicity is high on the plane, and becomes higher toward the galactic centre. On the other hand the [$\alpha$/Fe] ratios show a strong vertical gradient but no radial gradient except for the galactic bulge; the majority of the bulge stars have high [$\alpha$/Fe] ratios \citep{kob11mw}, which may depend on the feedback from the galactic centre.
There is no chemodynamical model that includes feedback from the central SMBH in the Milky Way.

\begin{figure}[t]
\centering
\includegraphics[width=0.65\textwidth]{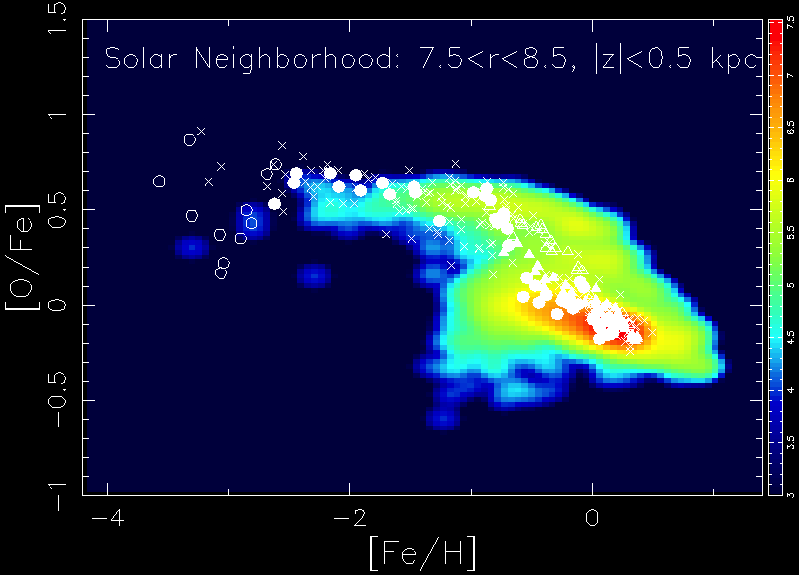}
\caption{The [O/Fe]--[Fe/H] relation in the solar neighborhood of our cosmological `zoom-in' simulation of a Milky Way-type galaxy at $z=0$. Figure is the same as Fig.\,10 of \citet{kob11mw} but with a newer simulation. The colour indicates the number of stars in logarithmic scale, showing a bimodality. The observational data sources are:
\citet[open and filled triangles]{ben04} for thick and thin disk stars, respectively,
\citet[open circles]{spi05},
\citet[filled circles]{zhao16}, 
\citet[crosses]{ama19b}.
}
\label{fig:mwofe}
\end{figure}

Figure \ref{fig:mwofe} shows the frequency distribution of [O/Fe] ratios in the solar neighborhood of our simulated galaxy.
The same [O/Fe]--[Fe/H] relation as in Fig.\,\ref{fig:ofe} is seen: the plateau at [O/Fe] $\sim 0.5$ caused by core-collapse supernovae, and the decrease of [O/Fe] from [Fe/H] $\sim -1$ to $\sim 0$ due to the delayed enrichment from SNe Ia.
The difference is that this chemodynamical model predicts not a line but a distribution (contours). A similar figure was also shown in \citet{kob11mw}, which predicted a bimodal distribution of [O/Fe] ratio at a given [Fe/H], concluding that ``this may be because the mixing of heavy elements among gas particles is not included in our model''. However, with the APOGEE survey \citep{hay15} clearly showed the bimodality of [$\alpha$/Fe], which was in fact already seen in earlier works with a much smaller sample but careful analysis \citep{fuhrmann98,ben03}. The advantage of the APOGEE survey was its large dynamic range and the change of the bimodality depending on the location within the Galaxy is also clearly shown. \citet{kob16iau} showed a similar change in the simulated Galaxy, and \citet{vin20} showed a comparison to APOGEE DR16. Our simulated galaxy predicts this bimodality not only for [$\alpha$/Fe] ratios but also for most of elements from He to U; some elements even show trimodality (\citealt{kob22iau}).

At [Fe/H] $\ltsim -1$, Figure \ref{fig:mwofe} also shows a significant number of stars with low [$\alpha$/Fe] ratios, which is caused by local enrichment from SNe Ia. Note that the SN Ia rate becomes almost zero from the stars below [Fe/H] $=-1.1$ in the adopted SN Ia model. Nonetheless, stars formed in less-dense regions can have low [$\alpha$/Fe] caused by SNe Ia, or by low-mass Type II supernovae (13--15$M_\odot$); this effect is important also for dSph galaxies.
The number of such low-$\alpha$ stars increases with sub-Ch mass SNe Ia \citep{kob20ia}.

The origin of the scatter/bimodality is discussed in \citet{kob14iau}. In chemodynamical simulations, it is possible to trace back the formation place of star particles. As a result, stars formed in merging galaxies found to have old age, low metallicity, high [$\alpha$/Fe] ratios, and relatively low [(Na,Al,Cu)/Fe] ratios, while stars formed in-situ show a tighter age-metallicity relation and low [$\alpha$/Fe] ratios. There is no age-metallicity relation for the stars that migrated, as expected.
In cosmological zoom-in simulations, stars tend to migrate outwards by a few kpc, which flatten the metallicity gradient by $\sim 0.05$ dex/kpc. Radial flows steepen the gradient as chemically enriched gas moves inwards, but the average velocity is found to be only $\sim 0.7$ km/s \citep{vin20}.

\subsubsection{The r-process sites}
Inhomogeneous enrichment in chemodynamical simulations leads to a paradigm shift on the chemical evolution of galaxies.
As in a real galaxy, i) the star formation history is not a simple function of radius, ii) the ISM is not homogeneous at any time, and iii) stars migrate, which are fundamentally different from one-zone or multi-zones GCE models.
As a consequence, (1) there is no tight age--metallicity relation, namely for stars formed in merging galaxies. It is possible to form extremely metal-poor stars at a later time, from accretion of nearly primordial gas, or in isolated chemically-primitive regions.
(2) Enrichment sources with long time-delays such as AGB stars, SNe Ia, and NSMs can appear at low metallicities.
(3) There is a significant scatter in elemental abundance ratios at a given time/metallicity, as shown in Figs.\,16--18 of \citet{kob11mw}.

\begin{figure}[t]
\begin{center}
  \includegraphics[width=0.95\textwidth]{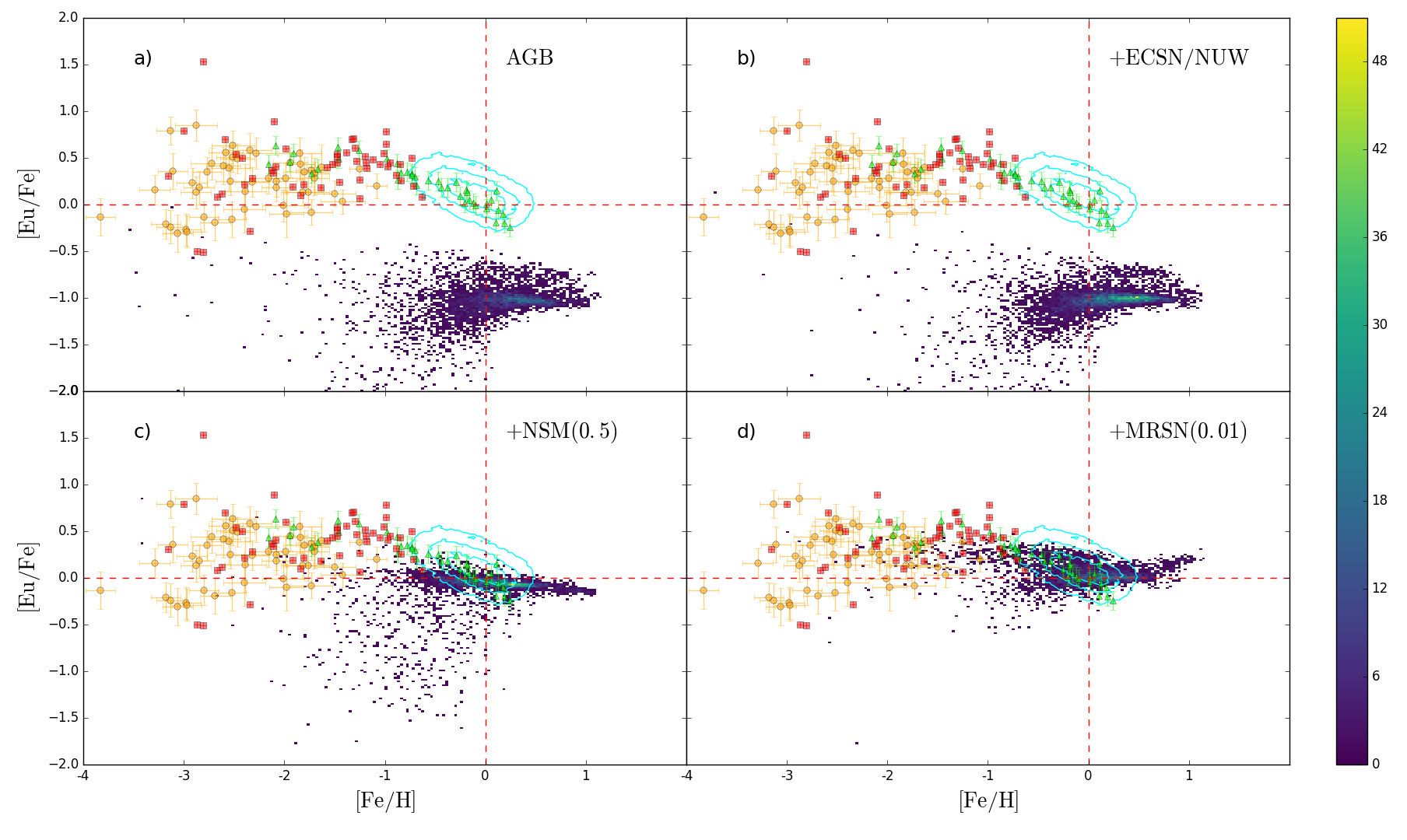}
\vspace*{-1mm}
\caption{Distribution of [Eu/Fe] against [Fe/H] for the star particles in the solar neighbourhood of our simulated Milky Way-type galaxy at $z=0$.
The panels in order show: control, ECSNe+$\nu$-driven winds, NSMs, and MRSNe. The observational data sources are: \citet[][red squares]{han14}, \citet[][orange circles]{roe14}, \citet[][green triangles]{zhao16}, and \citet[][GALAH DR2, cyan contours]{buder18}. The contours show 10, 50, and 100 stars per bin. The colour bar shows the linear number per bin of simulation data.
Figure is taken from \citet{hay19}.
}
\label{fig:mweu}
\end{center}
\end{figure}

Figure \ref{fig:mweu} shows the [Eu/Fe] ratios in the solar neighborhood of our simulated galaxy but with switching the r-process sites. With only AGB stars (panel a), Eu is not sufficiently produced. Different from one-zone GCE models, AGB contribution can be seen with a large scatter at low metallicity. 
With ECSNe or $\nu$-driven winds (panel b), Eu production is not increased enough. With NSMs (panel c), it is possible to reproduce the solar Eu/Fe ratios at the solar metallicity, but the scatter is too large at [Fe/H] $\ltsim 0$. Unlike one-zone GCE models, there are a small number of stars that have high [Eu/Fe] ratios at low metallicities due to the inhomogeneous enrichment.
The scatter can be much reduced with MRSNe (panel d) as they occur at a very short timescale. Since only a small fraction of core-collapse supernovae produce both Eu and Fe, the small scatter still remains, consistent with observations.

Very high [Eu/Fe] ratios were reported for ultrafaint dSphs \citep{ji16}, which might be caused by local enrichment with a NSM with no Fe production (and no Zn enhancement unlike \citet{yon21a}'s star). The effect of supernova kicks in binary neutron star systems may help \citep{van22}. Although supernova kicks are included in the delay-time distributions from BPSs, it is difficult to include it in the sub-galactic scale. In dSph galaxies, star formation takes place slowly, chemical enrichment proceeds inefficiently, and thus the inhomogeneous enrichment effect becomes even more important. Higher resolution of simulations with a new formalism/code will be required.

\subsection{Extra-galactic archaeology}
\label{sec:cosmo}

\begin{figure}[t]
\begin{center}
  \includegraphics[width=0.48\textwidth]{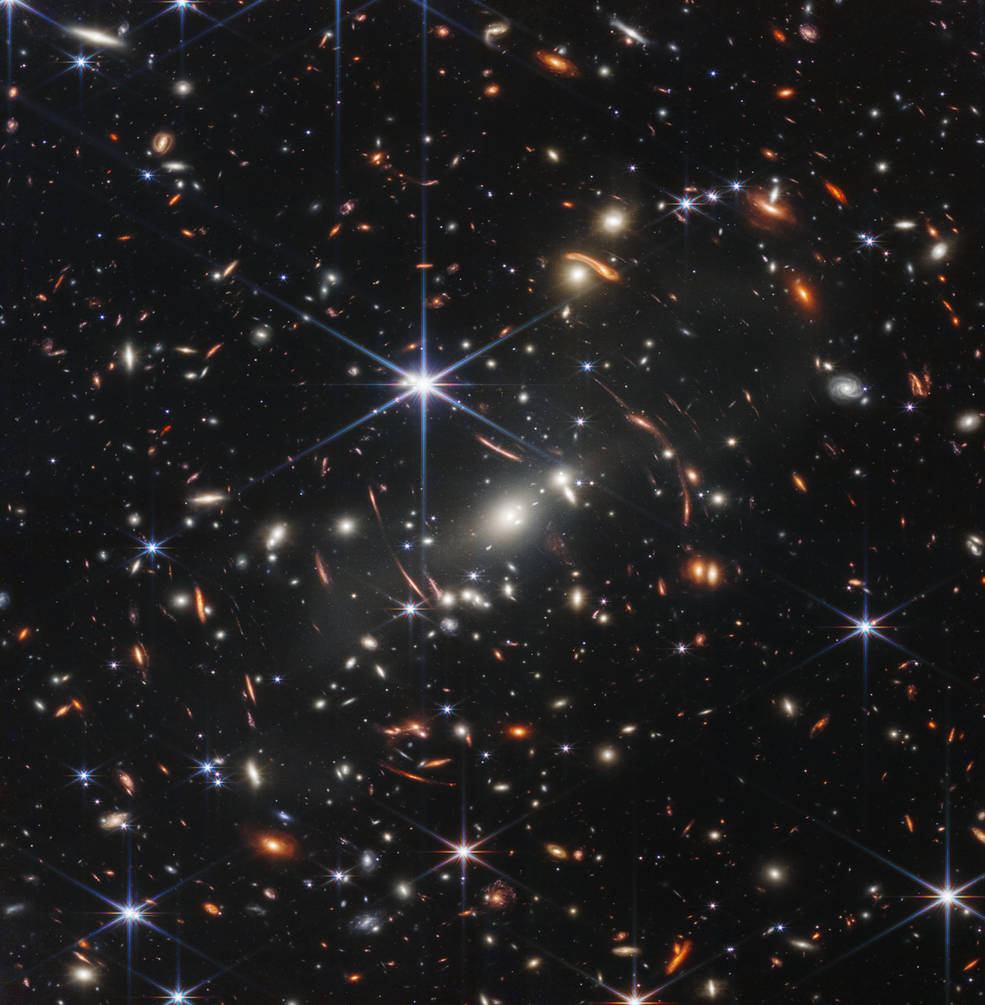}
  \includegraphics[width=0.5\textwidth]{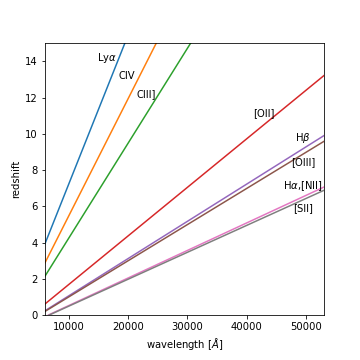}
\caption{(Left panel) The first NIRCam/JWST image of a cluster of galaxy SMAC0723 at $z=0.39$ but showing much higher redshift galaxies too, released on 11th July 2022. Figure is taken from \url{https://webbtelescope.org}.
(Right panel) Emission lines that can be obtained with NIRSpec/JWST as a function of wavelength and redshift.
}
\label{fig:jwst}
\end{center}
\end{figure}

Elemental abundances and isotopic ratios provide additional constrains on the timescales of formation and evolution of galaxies. In galaxies beyond the Local Group, it is not usually possible to resolve stars or HII regions, and metallicities of stellar populations\footnote{Multiple lines should be used to break the age-metallicity degeneracy.} or the ISM are estimated from absorption or emission lines in integrated spectra. Thanks to MOS surveys since SDSS the sample is greatly increased, and thanks to IFU surveys since SAURON the spatial (projected) distributions of metallicities are also obtained.
Note that the methods for obtaining the absolute values of physical quantities are still debated \citep{worthey92,conroy13,mai19,kew19}.
For stellar populations, $\alpha$/Fe ratios have been used to estimate the formation timescale of early-type galaxies \citep{tho05,kriek16}, while Fe is not accessible in star-forming galaxies, and instead CNO abundances can be used \citep{vin18a}. The JWST will push these observations toward higher redshifts, and the wavelength coverage of which will allow us to measure CNO abundance simultaneously (Figure \ref{fig:jwst}). More elements are available for X-ray hot gas or quasar absorption line systems, although neutron capture elements are out of reach.
Isotopic ratios of light elements and some light element abundances (e.g., F) are also estimated with ALMA (\S \ref{sec:firstgal}).

\begin{figure}[t]
\center
\url{https://star.herts.ac.uk/~chiaki/works/25mpc240_agn_gal0_zgas_1e-51e-1_opac3e13_1080p2.mp4}
\caption{Time evolution of gas-phase oxygen abundance around the most massive AGN-hosting galaxy at $z=0$ in a cosmological simulation \citep{tay15}.
The panel shows $10 \times 5.625$ Mpc$^2$ from the ($50h^{-1}$ Mpc)$^3$ cosmological simulation.
Colour shows the gas-phase oxygen abundance [O/H]=-5 (blue) to -1 (red) and $>-1$ (white).
[Credit: P. Taylor 2017]
}
\label{fig:cosmicmap}
\end{figure}

Using \citet{kob04}'s chemical enrichment routine, \citet{kob07} ran cosmological simulations with Gadget-3 and showed cosmic chemical enrichment.
\citet{tay14} included AGN feedback where BH seeds originate from the first stars, which results in earlier appearance of AGN feedback than other cosmological simulations.
The model parameters are chosen to match observational constraints: 1) the cosmic star formation rates \citep{mad14}, 2) the BH mass--velocity dispersion relation \citep{magorrian98,kor13}, 3) the size--mass relation of galaxies \citep[e.g.,][]{tru11}.
Although not used as constraints, simulated galaxies broadly agree with 4) the observed stellar mass function of galaxies, and 5) the star formation main sequence (SFMS).

Movie \ref{fig:cosmicmap} shows the evolution of gas-phase oxygen abundance around the most massive AGN-hosting galaxy at $z=0$ in a cosmological simulation \citep{tay15}.
Galaxies grow following the filamental structure of dark halos. Along the filaments, chemical enrichment is already on-going and supernova-driven winds eject metals from galaxies to the circum-galactic and inter-galactic medium. Metallicity is very inhomogeneous in space and time -- it can reach super-solar at the centre of galaxies at very high-redshifts while it stays primordial in the voids. 
Toward lower redshifts, SMBH grows in the central massive galaxy. This drives AGN-driven winds, ejecting remaining gas and metals and making this galaxy `red and dead'.

\begin{figure}[t]
\begin{center}
\includegraphics[width=0.70\textwidth]{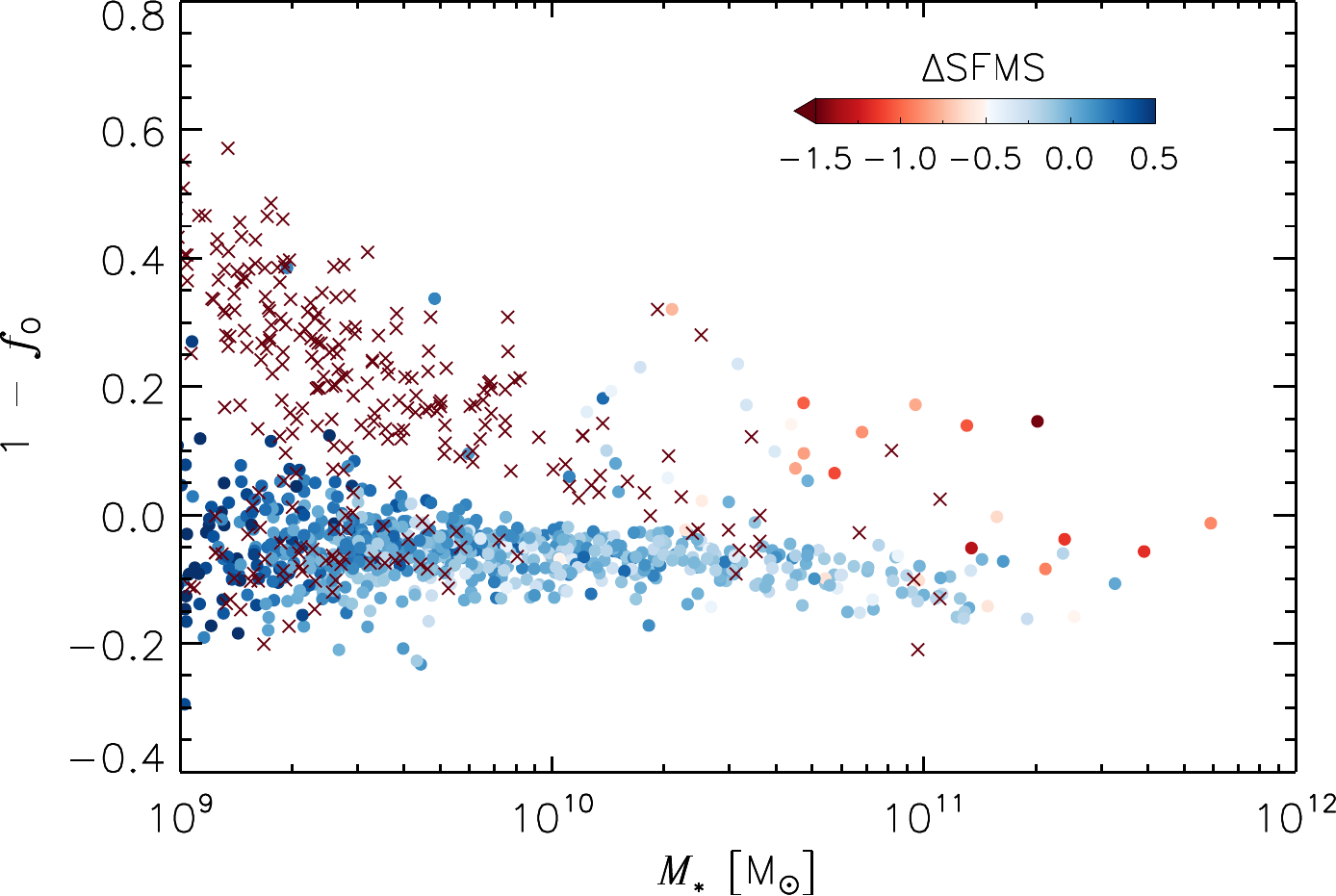}
\caption{\label{fig:zfms}
Mass fraction of oxygen lost from galaxies at $z= 0$ as a function of stellar mass of the galaxies in our cosmological simulation. Points are coloured by distance from the star formation main sequence (SFMS), and crosses denote galaxies with no star formation in the last $10^8$ yr.
Figure is taken from \citet{tay20}.
}
\end{center}
\end{figure}

\citet{tay20} analyzed the fraction of metals lost from galaxies, by using the quantities of present-day galaxies: stellar yields, initial stellar mass, and stellar metallicity and gas metallicity.
Figure \ref{fig:zfms} shows the mass fraction of oxygen lost from galaxies as a function of galaxy mass.
The galaxy population is bimodal; star-forming galaxies mostly occupy a tight sequence (which we refer to as the metal flow main sequence, or ZFMS) with a net gain of metals, whereas quenched galaxies tend to have lost metals.
Low-mass galaxies have lost up to $\sim$60\% of metals by supernova-driven winds, while massive galaxies can loose up to $\sim$20\% of metals by AGN-driven winds.
This trend is very similar to Fig.\,16 of \citet{kob07}, who used a different analysis method, and is explained due to the deeper potential well of massive galaxies.

\subsubsection{Mass--metallicity relations}
Both in observations and simulations, more massive galaxies tend to have higher metallicities, which is called {\it mass--metallicity relations} (MZR).
As already discussed, the origin of these relations are mass-dependent galactic winds by supernova feedback \citep{kob07,tay20}. 
Metal-loss by AGN feedback is not so important \citep{tay15letter}.
More intense star formation also happens in the simulated massive galaxies. An IMF variation is also possible \citep{kob10imf}. 

Without changing the IMF, it is possible to reproduce the observed MZRs at the present for stars and gas simultaneously \citep{tay15} . However, at higher redshifts, observational data shows a significant evolution, and gas-phase metallicities are a dex lower than observed. This problem is eased by changing the modelling of supernova feedback, such as to the mechanical feedback \citep{ibr24}. 
It is important to note that the method to estimate metallicities from emission line strengths is calibrated in the local Universe, and thus metallicities at higher redshifts are uncertain \cite{kew19}.

\subsubsection{Metallicity radial gradients}
Within galaxies, central parts are more metal-rich than outskirts, which is called {\it metallicity radial gradients}.
The origin of these gradients are the inside-out formation, where star formation starts from the centre with a higher efficiency and more chemical enrichment cycles.
The star formation is more intense at the centre in the simulated galaxies.
The star formation duration is not necessarily longer in the centre if quenching happens also inside-out by AGN feedback. 
Radial flows steepen the gradients, while galaxy mergers flatten the gradients \citep{kob04}.
Re-distribution of metals by AGN feedback may be important for gas-phase metallicity gradients of massive galaxies \citep{tay17}.

Simulations with the mechanical feedback can reproduce the metallicity gradients for stars and gas at the present \citep{ibr25}.
Gradients roughly correlate with the galaxy mass; more massive galaxies tend to have flatter gradients, although there is a significant scatter at a given mass \citep{kob04,tay17}.
This trend was explained by galaxy mergers in \citet{kob04}. This means that at higher redshifts, galaxies have not underwent mergers yet, and gradients become steeper.
This seems consistent with the observations of a few gravitational-lensed galaxies \citep{jones10,yuan11}.
However, recent IFU observations show a flat gradient around $z\sim2$ on the average \citep[e.g.,][]{curti20}, and some galaxies even show an inverse gradient \citep{cresci10}.
This is not reproduced even with changing supernova feedback modelling \citep{ibr25}.

\subsubsection{Elemental abundances}
Our cosmological simulations also predict elemental abundances and isotopic ratios.
At higher redshifts, it is harder to get absorption lines, and also galaxies are more star-forming, and thus \citet{vin18a} proposed to use CNO elements.
C is mainly produced from low-mass stars ($\ltsim 4M_\odot$), N is from intermediate-mass stars ($\gtsim 4M_\odot$) as a primary process \citep{kob11agb}, and O is from massive stars ($\gtsim 13M_\odot$). C and N are also produced by massive stars; the N yield depends on the metallicity as a secondary process, and can be greatly enhanced by stellar rotation (as for F).

In the nearby universe, the N/O--O/H relation is known for stellar and ISM abundances, which shows a plateau (N/O $\sim -1.6$) at low metallicities and a rapid increase toward higher metallicities. Damped Ly$\alpha$ systems at higher redshifts also roughly follow the same plateau. As discussed in \S \ref{sec:sn}, this relation was interpreted as the necessity of rotating massive stars by \citet{chi06}. However, this should be studied with hydrodynamical simulations including detailed chemical enrichment, and \citet{kob14iau} first showed the N/O--O/H relation in a chemodynamical simulation.
Figure \ref{fig:cno} shows a theoretical prediction on the time evolution of the N/O--O/H relation from \citet{vin18no}, where galaxies evolve along the relation. Recent observation with KMOS on VLT confirmed a near redshift-invariant N/O-O/H relation \citep[][KLEVER survey]{hay22}. 
However, recent observations with JWST found several galaxies far from this relation including GN-z11 (\ref{sec:firstgal}).

Figure \ref{fig:ofe-cosmic} shows [$\alpha$/Fe] ratios of stellar populations as a function of galaxy mass for our simulated galaxies with and without AGN feedback. Without AGN feedback (red diamonds), since star formation lasts longer in massive galaxies with a deeper potential well, [$\alpha$/Fe] ratios become lower in massive galaxies, which is the opposite compared with the observations (solid lines). With AGN feedback (black asterisks) star formation can be suppressed before the SN Ia enrichment becomes dominant in massive galaxies, so that [$\alpha$/Fe] ratios can stay high. However, the scatter of [$\alpha$/Fe] ratios at a given galaxy mass is still larger than observed.

[$\alpha$/Fe] ratios become higher at higher redshifts in the simulations, but not as much as observed \citep[e.g.,][]{kriek16}.
Some cosmological simulations claimed that they could reproduce the observed [$\alpha$/Fe]--mass relation at $z=0$, but with an ad-hoc SN Ia model. Some even introduced a variation in the IMF or in the binary fraction. 
However, the [$\alpha$/Fe] problem should be discussed with a chemodynamical model that is based on nuclear astrophysics and that can reproduce the observed elemental abundances both in the Milky Way and dSph galaxies. We do not have such a model yet.

\section{Conclusions and Future Prospects}

Thanks to the long-term collaborations between nuclear physics and astrophysics, we have good understanding on the origin of elements (except for the elements around Ti and a few neutron-capture elements such as Au; Fig.\,\ref{fig:origin}).
Elemental abundances, and isotopic ratios, are extremely useful for a number of questions in the Galactic and extragalactic archaeology: e.g., 1) the merging history of the Milky Way, 2) the origin of disks, 3) star formation histories of external galaxies, and 4) the IMF across cosmic time.
Although classic GCE models are useful for quickly checking the possible scenarios, chemodynamical simulations from cosmological initial conditions are required to confirm it.

\begin{figure}[t]
\begin{center}  \includegraphics[width=0.70\textwidth]{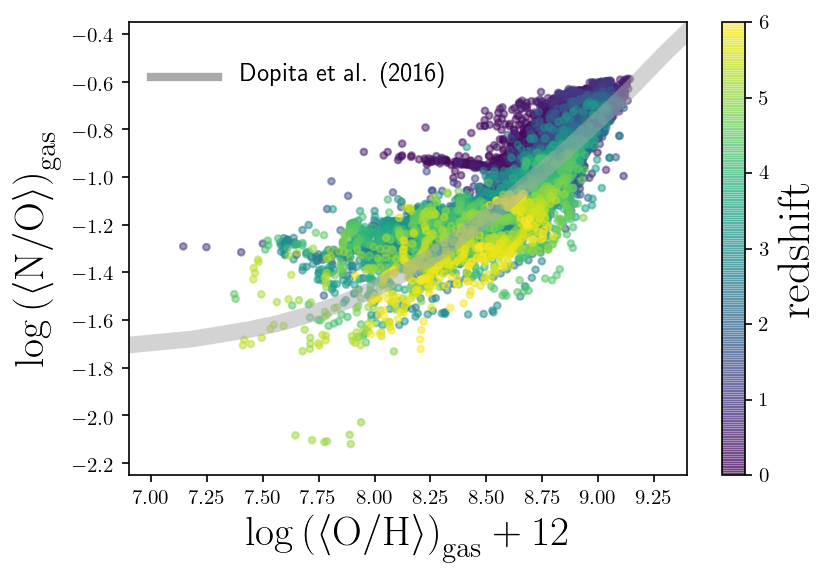}
\caption{Evolution of the gas-phase N/O--O/H relation of our simulated galaxies with halo masses $10^{11}M_\odot\le M_{\rm DM}\le 10^{13}M_\odot$ in a cosmological simulation. 
The abundances in the figure correspond to SFR-weighted averages in the gas phase of the ISM. The colour coding represents the redshift.
The gray bar indicates the compilation of observational data. 
Figure is taken from \cite{vin18no}.
}
\label{fig:cno}
\end{center}
\end{figure}

\begin{figure}[t]
\begin{center}
\vspace*{-4.cm}
\includegraphics[width=0.7\textwidth]{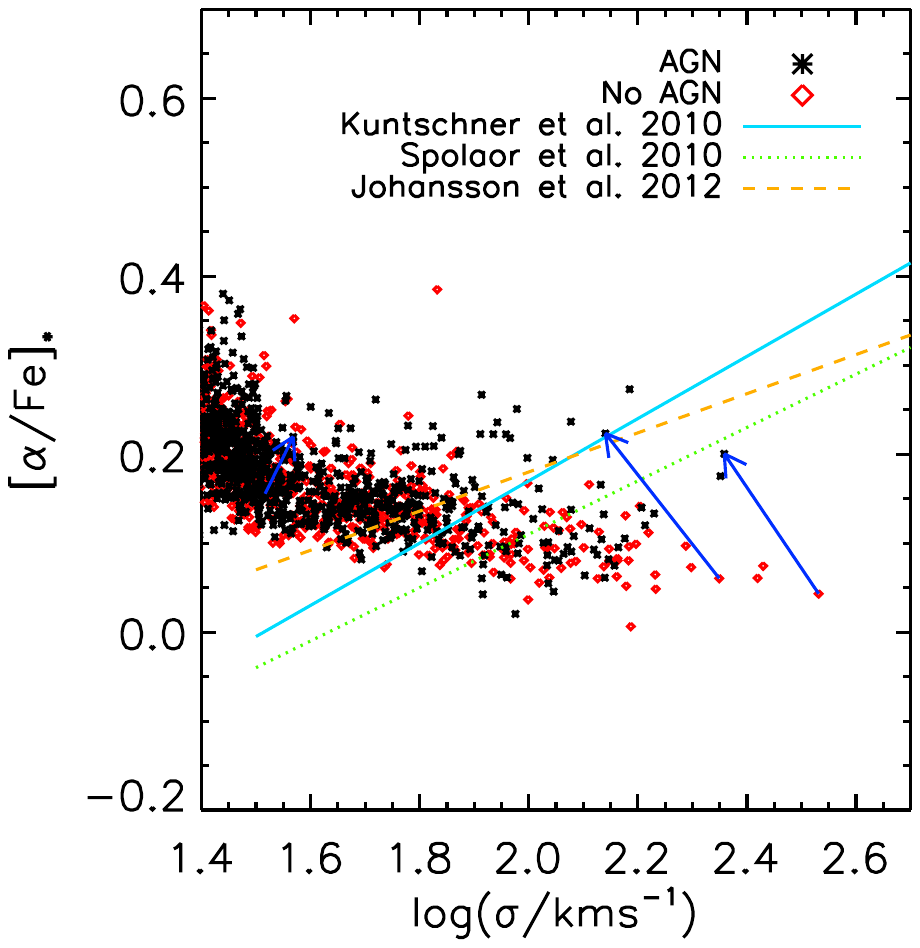}
\vspace*{-3.5cm}
\caption{Stellar [$\alpha$/Fe] ratios as a function of central velocity dispersion, which is a proxy of galaxy mass, for our simulated galaxies with (black asterisks) and without (red diamonds) AGN feedback at $z=0$. The arrows indicate the differences in two cosmological simulations from the same initial conditions. The solid lines indicate the observed relations.
Figure is taken from \citet{tay15}.
}
\label{fig:ofe-cosmic}
\end{center}
\end{figure}

In these simulations, inhomogeneous enrichment is extremely important for interpreting the elemental abundance trends.
The $\alpha$/Fe bimodality is naturally produced by the delayed enrichment from SNe Ia, even without a major merger (Fig.\,\ref{fig:mwofe}).
The observed N/O--O/H relation can be reproduced only with AGB stars and supernovae (Fig.\,\ref{fig:cno}).
However, the observed r-process abundances are not reproduced with NSMs alone (Fig.\,\ref{fig:mweu}); an r-process associated with core-collapse supernovae such as magneto-rotational hypernovae is required (Fig.\,\ref{fig:yong}), although the explosion mechanism is unknown.
It is necessary to run chemodynamical simulations  following detailed chemical enrichment `on-the-fly' based on nuclear astrophysics.

The impact of stellar rotation, binaries, and magnetic fields during stellar evolution, and the multi-dimensional effects of supernova explosions in nucleosynthesis should be investigated further.
Stellar rotation, or binary interaction, is required to reproduce WR stars in the local Universe. WR stars can explain the high F abundance of ALMA's high-redshift galaxy (Fig.\,\ref{fig:franco}), and also the high N abundance of JWST's very-high redshift galaxies \citep[e.g.,][]{bunker23}, together with intermittent star formation (Fig.\,\ref{fig:gnz11}).
Alternatively, the enrichment from very-massive stars (VMSs) or super-massive stars (SMSs) are considered; in the case of VMSs, the enrichment from PISNe needs to be avoided by somehow.

There are no stars enriched by a PISN in the Milky Way (e.g., high (Si,S)/O; Fig.\,\ref{fig:cdla}).
This means that no 140--300$M_\odot$ first stars were born in the Milky Way. Instead, the Milky Way halo was slowly enriched from 13--40$M_\odot$ faint supernovae (high C/Fe). It is still a mystery why stars explode so differently at $Z=0$. The signature of faint supernovae is also found in sub-DLAs \citep{saccardi23}. Where are the 140--300$M_\odot$ first stars then?

One might think that we should look at very distant galaxies launching an infrared telescope. However, if you know the theory of GCE, the chemical enrichment timescale is so short that it is very difficult to find the relics of the first stars in galaxies.
The outskirts of galaxies or primordial pockets in the nearby Universe might be an alternative target, if massive stars can form in such low-density regions.
Ideally, we would like to see high-resolution spectra of low-mass star-forming galaxies in the early Universe, and thankfully gravitational lensing in deep JWST imaging data provided a small number of candidates at fairly high redshifts \citep[e.g.,][]{vanzella23}.

\begin{ack}[Acknowledgments]

We thank P. Taylor, F. Vincenzo, D. Yong, A. Karakas, M. Lugaro, N. Tominaga, S.-C. Leung, M. Ishigaki, and K. Nomoto,
for fruitful discussion, and V. Springel for providing Gadget-3.
CK acknowledge funding from the UK Science and Technology Facility Council through grant ST/M000958/1, ST/R000905/1, ST/V000632/1, ST/Y001443/1.
The work was also funded by a Leverhulme Trust Research Project Grant on ``Birth of Elements''.
This work used the DiRAC Memory Intensive service (Cosma8 / Cosma7 / Cosma6 [*]) at Durham University, managed by the Institute for Computational Cosmology on behalf of the STFC DiRAC HPC Facility (www.dirac.ac.uk). The DiRAC service at Durham was funded by BEIS, UKRI and STFC capital funding, Durham University and STFC operations grants. DiRAC is part of the UKRI Digital Research Infrastructure.

\end{ack}

\bibliography{book}{}
\bibliographystyle{aasjournal}

\end{document}